\documentclass[prd,showpacs,aps,amsmath,amssymb,nofootinbib,floatfix,floats,superscriptaddress,twocolumn]{revtex4-1} 

\allowdisplaybreaks[2]

\usepackage[usenames, dvipsnames]{color}
\usepackage[colorlinks, pdfborder={0 0 0}, plainpages=false]{hyperref}
\usepackage{amsmath, amssymb, amsfonts}
\usepackage{acronym}
\usepackage{xspace} % Sensible space treatment at end of simple macros
\usepackage{dcolumn}% Align table columns on decimal point
\usepackage{bm}% bold math
\usepackage{multirow} % Multiple rows in tables
\usepackage[utf8]{inputenc} % for some references from inspirehep.net
\usepackage{graphicx}
\usepackage{longtable}
\usepackage{amsmath}
\usepackage{times}
\usepackage{mathptmx}
\usepackage{epstopdf}
\usepackage{textcomp}

\def\laq{\raise 0.4ex\hbox{$<$}\kern -0.8em\lower 0.62ex\hbox{$\sim$}}
\def\gaq{\raise 0.4ex\hbox{$>$}\kern -0.7em\lower 0.62ex\hbox{$\sim$}}

\newcommand{\AEI}{\affiliation{Max Planck Institute for Gravitational Physics (Albert Einstein Institute), Am M\"uhlenberg 1, Potsdam-Golm, 14476, Germany}}
\newcommand{\Maryland}{\affiliation{Department of Physics,University of Maryland, College Park, MD 20742, USA}}
\newcommand{\Cardiff}{\affiliation{School of Physics and Astronomy, Cardiff University, Cardiff CF24 3AA, Wales, UK}}

\def\be{\begin{equation}}
\def\ee{\end{equation}}
\def\bea{\begin{eqnarray}}
\def\eea{\end{eqnarray}}
\newcommand{\bes}{\begin{subequations}}
\newcommand{\ees}{\end{subequations}}

\begin{document}

\title{Black-hole Spectroscopy by Making Full Use of Gravitational-Wave Modeling}

\author{Richard Brito}\thanks{richard.brito@aei.mpg.de} \AEI 
\author{Alessandra Buonanno}\thanks{alessandra.buonanno@aei.mpg.de} \AEI \Maryland 
\author{Vivien Raymond}\thanks{raymondv@cardiff.ac.uk} \AEI \Cardiff
\date{\today}

\begin{abstract}
The Kerr nature of a compact-object--coalescence remnant can be unveiled by observing multiple quasi-normal modes in the post-merger signal. Current methods to achieve this goal rely on matching the data with a superposition of exponentially damped sinusoids with amplitudes fitted to numerical-relativity (NR) simulations of binary black-hole mergers. These models presume the ability to correctly estimate the time at which the gravitational-wave signal starts to be dominated by the quasi-normal modes of a perturbed black hole. Here we show that this difficulty can be overcome by using multipolar inspiral-merger-ringdown waveforms,  calibrated to NR simulations, as already developed  within the effective-one-body formalism (EOBNR). We build a parameterized (nonspinning) EOBNR waveform model in which the quasi-normal mode complex frequencies are free parameters (pEOBNR), and use Bayesian analysis to study its effectiveness  in measuring quasi-normal modes in GW150914, and in synthetic gravitational-wave signals of binary black holes injected in Gaussian noise. We find that using the pEOBNR model gives, in general, stronger constraints compared to the ones obtained when using a sum of damped sinusoids and using Bayesian model selection, we also show that the pEOBNR model can successfully be employed to find evidence for deviations from General Relativity in the ringdown signal. Since the pEOBNR model properly includes time and phase shifts among quasi-normal modes, it is also well suited to consistently combine information from several observations --- e.g., we find on the order of $\sim 30$ GW150914-like binary black-hole events would be needed for Advanced LIGO and Virgo at design sensitivity to measure the fundamental frequencies of both the $(2,2)$ and $(3,3)$ modes, and the decay time of the $(2,2)$ mode with an accuracy of $\lesssim 5\%$ at the $2\mbox{-}\sigma$ level, thus allowing to test the black hole's no-hair conjecture.
\end{abstract}
\maketitle

\acrodef{PDF}[PDF]{probability density function}
\acrodef{PSD}[PSD]{power spectral density}
\acrodef{LAL}[LAL]{LIGO Algorithm Library}

%{\bf Introduction.} 
\section{Introduction}
Up to now, all the observed gravitational waves (GWs) from the
coalescence of compact objects by Advanced LIGO and
Virgo~\cite{Abbott:2016blz,
  Abbott:2016nmj,Abbott:2017vtc,Abbott:2017gyy,Abbott:2017oio}
are entirely consistent with the expected gravitational radiation
emitted during the inspiral, merger and ringdown stages of a binary black hole (BBH), as predicted by Einstein theory of General
Relativity
(GR)~\cite{TheLIGOScientific:2016src,TheLIGOScientific:2016pea} (however, note that GW170817~\cite{TheLIGOScientific:2017qsa,Abbott:2018wiz} was most likely a binary neutron star event due to its electromagnetic counterpart). After
merger, GR predicts that the remnant black hole (BH) is described by
the Kerr metric~\cite{Kerr:1963ud}, the unique stationary,
axisymmetric and asymptotically flat BH solution of the Einstein field
equations in vacuum (astrophysical BHs are thought to be
electrically neutral). As detectors with improved sensitivity and
longer observation times come online, the signal-to-noise ratio (SNR)
and number of events will increase, and more stringent gravitational
tests could put GR at stake~\cite{Will:2014kxa,Berti:2015itd}, and/or
reveal the existence of exotic astrophysical compact
objects~\cite{Mazur:2004fk,Visser:2003ge,Liebling:2012fv} in our
Universe.

Consistent with theoretical predictions, the GW signals of the five
BBHs observed so far by Advanced LIGO, GW150914, GW151226, GW170104, GW170608
and GW170814, chirp from the inspiral stage, where the orbital
frequency increases as the two objects come closer and closer, up to
merger, where the GW luminosity reaches a peak and non-perturbative GR
effects dominate. After the merger, the waveform settles to a linear
superposition of exponentially damped sinusoidal oscillations
(ringdown) or quasi-normal modes (QNMs), described by a discrete set
of complex frequencies which are uniquely determined by the nature of
the remnant BH and are independent on how the BH was formed. That BHs,
when formed and/or perturbed, emit GWs described by a very
specific set of QNMs was discovered in the early
70s~\cite{Vishveshwara:1970zz,Press:1971wr,Chandrasekhar:1975zza}. In
vacuum, the no-hair theorems~\cite{Israel:1967wq,Carter:1971zc,Hawking:1971vc,Robinson:1975bv}
imply that in GR the BH's QNMs depend only on the BH's mass $M_{\rm BH}$ and
angular momentum (or spin) $J_{\rm BH}$, and therefore testing this hypothesis
requires the identification of at least two QNMs in the
ringdown waveform~\cite{Dreyer:2003bv,Berti:2005ys,Gossan:2011ha,Meidam:2014jpa,Yang:2017zxs,DaSilvaCosta:2017njq}
~\footnote{Several examples within GR that do not satisfy the conditions of the no-hair conjecture have been constructed. However, most of those solutions either lead to instabilities or they require the presence of exotic fields or time-dependent boundary conditions for complex boson fields  (see,  e.g., Ref.~\cite{Cardoso:2016ryw}).}.

The idea of employing spectroscopy of the ringdown stage of
compact-object binary mergers to prove that a BH has been observed (or
better rule out/constrain theories alternative to GR or other compact
objects rather than BHs) and test the no-hair hypothesis in GR, was
first examined in Ref.~\cite{Dreyer:2003bv}. Later,
Ref.~\cite{Berti:2005ys} carried out a comprehensive study aimed at
quantifying the accuracy with which the QNM (complex) frequencies can
be measured for GW sources observable by the laser-interferometer
space-based antenna (LISA), and applied statistical criteria to
estimate the resolvability of different modes. The latter was also
used in subsequent publications (e.g., see
Refs.~\cite{Berti:2007zu,Bhagwat:2016ntk,Berti:2016lat,Maselli:2017kvl}), which
focused also on future GW detectors on the ground. An important step
in understanding the feasibility of the BH--spectroscopy program came
with Refs.~\cite{Gossan:2011ha,Meidam:2014jpa}, where the authors
applied Bayesian techniques for the first time, employed parameterized
models for the ringdown signals and advocated for the use of multiple
events to get stronger tests of the GR no-hair conjecture. More
recently, Ref.~\cite{Yang:2017zxs} proposed a strategy to increase the 
accuracy of observing a given QNM by constructively summing
the ringdown signal from multiple events, after appropriately applying 
a rescale and time shift such that the QNM in all signals has the same
frequency and phase. The same idea was proposed in
Ref.~\cite{DaSilvaCosta:2017njq} although only implemented for the
least-damped QNM. We stress, that this recent idea to extract subdominant modes 
relies on using the measured BBH parameters (i.e., masses and spins), and importantly 
on knowing in advance the relative phases and amplitudes of the excited QNMs.

In previous analyses of the BH--spectroscopy program, all studies were
conducted employing for the ringdown signal a superposition of exponentially 
damped sinusoids with either free amplitudes and phases~\cite{Berti:2005ys,Maselli:2017kvl,Cabero:2017avf}
or with amplitudes fitted to numerical-relativity (NR) simulations~\cite{Berti:2007zu,Gossan:2011ha,Meidam:2014jpa}. Here, by contrast,
we make {\it full} use of GW modeling from BBH coalescences and employ
inspiral-merger-ringdown (IMR) waveforms as developed within the
effective-one-body formalism~\cite{Buonanno:1998gg,Buonanno:2000ef}, augmented by 
NR simulations~\cite{Pan:2011gk} (EOBNR waveforms,
for short~\footnote{The specific name of the waveform
  model that we use in the \textsc{LIGO Algorithm Library} is
  \textsc{EOBNRv2HM}.}). There are two main advantages in doing so. First, EOBNR
waveforms include, by construction, the phase difference between
different QNMs, tuned to NR simulations, thus avoiding to apply sophisticated techniques to
enforce such a coherence a posteriori (i.e., after the
observation~\cite{Yang:2017zxs,DaSilvaCosta:2017njq}). Second, there is no need to define an
\emph{a priori} unknown time at which the QNMs start to dominate the
post-merger signal (or select a few arbitrary values, as was done for GW150914 \cite{TheLIGOScientific:2016src}), because this time is automatically taken into
account when building EOBNR waveforms, so that they match NR waveforms
with high precision. As we shall see, the apparent limit in the
accuracy of extracting QNM frequencies, as recently advocated in
Ref.~\cite{Thrane:2017lqn}, does not hold when employing IMR
waveforms.
     
The rest of this paper is organized as follows. We first introduce our
IMR waveform model with free QNM complex frequencies in
Sec.~\ref{sec:model}, and discuss how this model can be used to measure
the ringdown frequencies and damping time of a BBH-coalescence remnant. 
In Sec.~\ref{sec:PE} we present the statistical method that we
employ to measure the QNM complex frequencies, and test the IMR model
against the GW event GW150914 and NR waveforms. Section~\ref{sec:no_hair}
studies two different approaches to measure deviations from GR using
the IMR waveform model. We first perform a Bayesian model selection
study to show that the IMR model is able to find evidence for
deviations from GR in the ringdown. Then, we give some prospects, using
Advanced LIGO and Virgo noise curves at design sensitivity, on how
strongly the model can constrain deviations from GR by combining
several detections. Finally, we summarize and discuss future
improvements in Sec.~\ref{sec:conclusions}.

\section{Full gravitational-wave signal to extract quasi-normal modes}\label{sec:model}

We use the IMR waveforms developed within the EOB formalism, which provides  
a faithful and physical, semi-analytic description of the full coalescence process, and it can be made highly
 accurate by including information from NR simulations. In particular, here we employ 
the multipolar waveform model for nonspinning BBHs calibrated 
to NR simulations in Ref.~\cite{Pan:2011gk} (henceforth, EOBNR for short). 
A GW emitted from a binary into a given sky direction $(\theta, \phi)$ can be written as 
$h_+(\theta,\phi;t ) - i h_\times(\theta,\phi;t) = \sum_{\ell, m} {}_{-\!2}Y_{\ell m}(\theta,\phi)\, h_{\ell m}(t)$, 
where ${}_{-\!2}Y_{\ell m}(\theta,\phi)$ are the $-2$ spin-weighted spherical harmonics. 
Our EOBNR model includes the $(\ell, |m|)=(2,1)$, $(3,3)$, $(4,4)$, and $(5,5)$ modes besides the dominant $(2,2)$ mode.  

More specifically, for each $(\ell, m)$, the merger-ringdown EOBNR modes read
\begin{equation}
  \label{RD}
  h_{\ell m}^\mathrm{merger-RD}(t) = \sum_{n=0}^{N-1} A_{\ell mn}\,
  e^{-i\sigma_{\ell mn} (t-t_\mathrm{match}^{\ell m})} \quad t \geq t_{\rm match}^{\ell m}\,,
\end{equation}
where $n$ is the QNM overtone number, $N$ is the number of overtones included in the EOBNR model (e.g., 
$N = 8$ in Ref.~\cite{Pan:2011gk}~\footnote{We note that some of the high overtones used 
in Ref.~\cite{Pan:2011gk} do not have the frequency and 
decay time of a BH, and they were included {\it only} to make the merger-ringdown transition as 
smooth as possible.}), and $A_{\ell  mn}$ are complex amplitudes determined by the procedure 
that matches the merger-ringdown waveform to the inspiral-plunge EOBNR 
waveform $h_{\ell m}^\mathrm{inspiral-plunge}(t)$. Such a procedure guarantees differentiability at the matching 
point $t_\mathrm{match}^{\ell m}$\footnote{We note that making $t_\mathrm{match}^{\ell m}$ independent of the overtone number
was found to be enough to match the NR waveforms very well in Ref.~\cite{Pan:2011gk}.}. The quantity $\sigma_{\ell mn} = \omega_{\ell m n} -i/\tau_{\ell m n}$, where the oscillation frequencies $\omega_{\ell m
  n}>0$ and the decay times $\tau_{\ell m n}>0$, are numbers
associated with each QNM.  It was found in Refs.~\cite{Barausse:2011kb,Pan:2011gk}, that in the test-particle limit 
and comparable-mass case, the different modes can peak at different times, depending on mass ratio and spin values. 
We stress that the multipolar EOBNR model adopted here {\it does} reproduce this important feature by including appropriate 
time shifts between the modes ($\Delta^{\ell m}_{\rm match}$) in the matching procedure (for details see Fig. 1 and Sec. IIB in Ref.~\cite{Pan:2011gk}). Through this work we will use the same time shifts, $\Delta^{\ell m}_{\rm match}$, obtained in Ref.~\cite{Pan:2011gk}. Finally, the inspiral-(plunge-)merger-ringdown EOBNR waveform reads
$h_{\ell m}(t) = h_{\ell m}^\mathrm{insp-plunge}\, \theta(t_\mathrm{
    match}^{\ell m} - t) + h_{\ell m}^\mathrm{merger-RD}\,\theta(t-t_\mathrm{
    match}^{\ell m})$, where $\theta(t)$ is the Heaviside step function.

%%%%%%%%%%%%%%%%%%%%%%%%%%%%%%%%%%%%%%%%%
\begin{figure}[t]
\begin{center}
\includegraphics[width=0.48\textwidth]{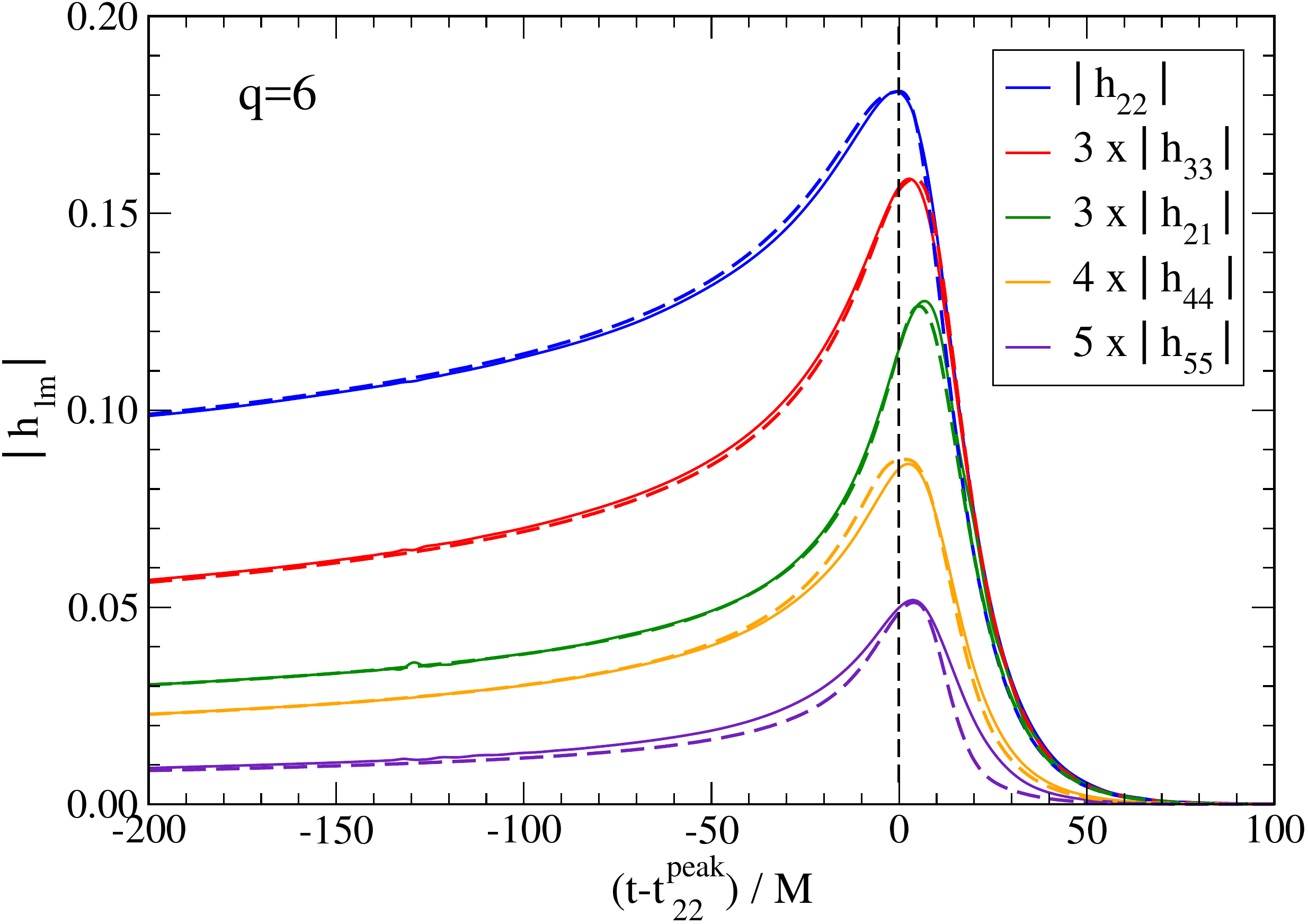}
\caption{Comparison between modes' amplitudes of the EOBNR model~\cite{Pan:2011gk} used here (dashed lines) and the NR waveform (solid lines) for a BBH 
simulation with mass ratio $q=6$ produced by the SXS collaboration~\cite{Mroue:2013xna}. In the horizontal axis 
the time origin is chosen such that it corresponds to the peak of the  $(2,2)$ mode.
  \label{fig:EOBvsNR}}
\end{center}
\end{figure}
%%%%%%%%%%%%%%%%%%%%%%%%%%%%%%%%%%%%%%%%%

  In Ref.~\cite{Pan:2011gk}, the complex frequencies $\sigma_{\ell m}$
  were expressed in terms of the final BH mass and
  spin~\cite{Berti:2005ys}, and the latter were related to the BBH's
  component masses and spins through an
  NR--fitting-formula~\cite{Pan:2011gk} computed in GR. For
  concreteness, in Fig.~\ref{fig:EOBvsNR} we show an example where we
  compare the amplitude of the different modes available in the EOBNR
  waveform, for a BBH with mass ratio $q=6$, against the
  waveform obtained from a NR simulation~\footnote{The NR waveforms
    used in this paper are from the Simulating eXtreme Spacetimes
    (SXS) catalog in Ref.~\cite{Mroue:2013xna}. The modes' amplitudes shown 
    in Fig.\ref{fig:EOBvsNR} refer to SXS:BBH:0166.}. Importantly, the model includes time shifts
  between the peak of each mode and agrees very well with NR, even for
  $\ell>2$-modes.

  Here, to measure the ringdown frequencies and damping times of
  different QNMs, we build a parameterized EOBNR model by relaxing the assumption that the ringdown signal is
  fixed by the NR--fitting-formula in Ref.~\cite{Pan:2011gk}, and
  instead promote the QNM (complex) frequencies to be free
  parameters (henceforth, pEOBNR model). In the specific applications of this paper, we will only
  allow $\sigma_{220}$ and $\sigma_{330}$ to vary freely, while
  all the other mode frequencies present in the merger-ringdown
  waveform coincide to the GR values. We emphasize that $\sigma_{220}$
  and $\sigma_{330}$ varying freely implies that the EOBNR waveform at
  merger (i.e., close to the peak and at $t_\mathrm{match}^{\ell m}$),
  {\it does not necessarily} coincide with the GR prediction, since the
  matching procedure changes the shape of the waveform for $t>
  t_\mathrm{match}^{\ell m}$ for $(\ell,m) = (2,2)$ and $(3,3)$.
  Lastly, for $t < t_{\rm match}^{\ell m}$, our EOBNR waveform modes
  agree with the inspiral-plunge modes $h_{\ell
    m}^\mathrm{inspiral-plunge}(t)$ computed in GR. In the future, as
  the EOB formalism is extended to modified theories of GR~\cite{Julie:2017pkb,Julie:2017ucp}, 
we will include non-GR inspiral-plunge modes and other possible variations 
around merger.

In the following, we contrast the results obtained with the pEOBNR
model, with a waveform model that consists of solely a superposition
of damped sinusoids, whose (complex) frequencies are free
parameters~\cite{Dreyer:2003bv,Berti:2005ys}. This has been the most
common ringdown model used in the literature to test the no-hair
conjecture and/or extract multiple QNMs. After the NR breakthrough in
2005, the relative amplitudes and phases of the QNMs in these models
have been constrained using fits from NR simulations of
BBHs~\cite{Kamaretsos:2011um,Gossan:2011ha,London:2014cma,London:2018gaq}.
More explicitly, the ringdown model that we employ is ($t \geq
0$)
\begin{align}
 h^{\rm RD}_{+}(\theta,\phi;t) = \sum_{\ell,m > 0} A_{\ell |m|}\, e^{-t/\tau_{\ell m}} \,Y^{\ell m}_{+}(\theta) \cos(\omega_{\ell m}t - \phi_{\ell m})\,, 
\label{eq:rd1} \\
h^{\rm RD}_{\times}(\theta,\phi;t) = - \sum_{\ell,m > 0} A_{\ell |m|}\, e^{-t/\tau_{\ell m}} \,Y^{\ell m}_{\times}(\theta) \sin(\omega_{\ell m}t - \phi_{\ell m}),
\label{eq:rd2}
\end{align}
where $Y_+^{lm}\equiv{}_{-2}Y^{lm}+(-1)^l~ _{-2}Y^{l-m}$ and $Y_\times^{lm}\equiv{}_{-2}Y^{lm}-(-1)^l~ _{-2}Y^{l-m}$, 
and $h_+=h_\times=0,$ for $t<0$, $t=0$ being the starting time of the ringdown signal. Since we focus on nonspinning BBHs, we use for the relative modes' amplitudes the NR-fits in Ref.~\cite{Gossan:2011ha}, so that the only free parameters are the mode frequencies $\omega_{\ell m}$, damping times $\tau_{\ell m}$, the phases $\phi_{\ell m}$, the BBH mass ratio $q$ and an overall amplitude factor (see Eqs. (5)--(8) in Ref.~\cite{Gossan:2011ha}). One crucial difference of this ringdown model from the pEOBNR model discussed above, is that the former assumes that all modes start at the same time, and this is not observed in NR simulations of BBHs (see Fig.~\ref{fig:EOBvsNR} and Ref.~\cite{Pan:2011gk}). Furthermore, 
the pEOBNR model also includes overtones beyond $n=0$, which can be excited around merger, as also observed in NR 
simulations~\cite{Buonanno:2006ui,Berti:2007fi}.

An important difficulty to overcome when using a damped sinusoid model is the 
need to define a specific starting time at which
the GW signal is well described by a sum of QNMs. Since the arrival time of the signal 
in the different detectors is a function of the sky position, 
to correctly define the time at which the ringdown starts 
in all detectors, one not only needs to know the geocentric time at coalescence 
but also the sky position of the signal~\cite{Cabero:2017avf}.
For a real event these parameters are \emph{a priori} unknown and must be obtained from a
previous analysis done with an IMR waveform. In addition to this
difficulty, to avoid biases and accurately recover the ringdown
parameters for an IMR signal, we also find it necessary to zero
out the synthetic GW signals injected in Gaussian noise prior to the starting time of the damped
sinusoid model. This behavior was already pointed out in
Ref.~\cite{Cabero:2017avf}, and is related to matching a model
with a cutoff in the time domain to a signal that includes all the IMR
information. These technical difficulties can be completely avoided by
using an IMR model, and therefore provide an additional motivation
for this work.

In summary, focusing on nonspinning BBHs with component masses $m_1$ and $m_2$, we consider two different waveform models: (i)
the pEOBNR waveform built from Ref.~\cite{Pan:2011gk} with free
parameters $\boldsymbol{\vartheta}_{\rm GR}=\{M_c,q,
D_L,\alpha,\delta,\psi,\theta,t_c, \phi_c\}$, where $M_c= M\nu^{3/2}$ is the
(redshifted) chirp mass, with $\nu = m_1m_2/(m_1+m_2)^2$ and 
$M$ the (redshifted) total mass, $q=m_1/m_2>1$ is the mass ratio, $D_L$ is the luminosity
distance, $\theta$  is the inclination angle of the binary, $\alpha$,
$\delta$ and $\psi$ are the right ascension, declination and
polarization angles, respectively, and $t_c$ and $\phi_c$ are the
(geocentric) time and phase at coalescence, supplemented with free
complex QNM frequencies for the (220) and (330) modes
$\boldsymbol{\vartheta}=\boldsymbol{\vartheta}_{\rm GR} \cup
\{\omega_{220},\tau_{220},\omega_{330},\tau_{330}\}$; and (ii) the damped
sinusoid model given by Eqs.~\eqref{eq:rd1} and (\ref{eq:rd2}). In this work we either
use only one damped sinusoid, or use a two-damped sinusoid model with
relative amplitudes for the (220) and (330) modes fitted to NR as
given in Ref.~\cite{Gossan:2011ha}, neglecting all the other
modes. Therefore for the two-damped sinusoid model the free parameters
are $\boldsymbol{\vartheta}_{\rm
  RD}=\{\omega_{220},\omega_{330},\tau_{220},\phi_{220},\tau_{330},
\phi_{330}, A, q\}$, with $A$ an overall amplitude, that can be
related to the BH final mass and the luminosity distance, while for
the single damped sinusoid model, the free parameters are simply
$\boldsymbol{\vartheta}_{\rm RD}=\{\omega_{220},\tau_{220},\phi_{220},
A\}$. We note that for both sinusoid models we fix the sky location $\{\alpha,\delta\}$ and geocentric time at coalescence $t_c$ which can be obtained by first performing parameter estimation using an IMR model. The damped sinusoid model is then chosen to start at a given fixed time after the coalescence time such as to fit only the ringdown part of the signal.

\section{Inference with the parameterized inspiral-merger-ringdown model}\label{sec:PE}
We now use Bayesian analysis~\cite{Bayes:1793, Jaynes:2003} to test the ability 
of the pEOBNR model to recover the QNM complex frequencies. In particular, we infer the ringdown-signal's 
parameters of GW150914~\cite{Abbott:2016blz}, which, so far, is the loudest BBH event detected 
by Advanced LIGO, and the only event with a non-negligible amount of SNR in the ringdown, and of a few synthetic GW 
signals injected in Gaussian noise. For the latter we employ two nonspinning NR waveforms from the SXS catalog~\cite{Mroue:2013xna}: 
(i) one with mass ratio $q=1.5$ (SXS:BBH:0007) and total mass $M=70 M_\odot$, which mimics the GW150914
event, and (ii) another with mass ratio $q=6$ (SXS:BBH:0166) and total mass $M=84 M_\odot$, 
for which modes with $l>2$ are non-negligible --- e.g., at merger the $(3,3)$-mode is $\sim 70\%$ smaller  
than the dominant $(2,2)$-mode in the face-on/off binary configuration (see Fig.~\ref{fig:EOBvsNR}).

We estimate the \ac{PDF} for a parameter vector 
$\vartheta$ according to the \textsc{LIGO Algorithm Library}
sampling algorithm in Ref.~\cite{veitch:2014wba}. We sample the posterior
density $p(\boldsymbol{\vartheta}|h, d)$ for the model $h$ given the data $d$ as a function
of $\boldsymbol{\vartheta}$ using:
\begin{equation}
p(\boldsymbol{\vartheta}|h, d) \propto 
\mathcal{L}(d|\boldsymbol{\vartheta}) \times p(\boldsymbol{\vartheta})\,,
\label{eq:bayes}
\end{equation}
where $\mathcal{L}(d|\boldsymbol{\vartheta})$ is the
likelihood function of the observed data for given values of the
parameters $\boldsymbol{\vartheta}$, and $p(\boldsymbol{\vartheta})$
is the prior probability density of the unknown parameter vector
$\boldsymbol{\vartheta}$. To obtain the likelihood function
$\mathcal{L}(d|\boldsymbol{\vartheta})$, we first generate
the GW polarizations $h_+(\boldsymbol{\vartheta})$
and $h_{\times}(\boldsymbol{\vartheta})$ according
to the waveform models described above. We then combine the
polarizations into the two Advanced LIGO and Advanced Virgo detector responses at design
sensitivity, $h_{1,2,3}$, by projecting them on the detector antenna
patterns~\cite{Finn:1992wt}: $h_k(\boldsymbol{\vartheta}) = h^k_{+}(\boldsymbol{\vartheta})\,F^{(+)}_k(\boldsymbol{\vartheta}) 
+h^k_{\times}(\boldsymbol{\vartheta})\,F^{(\times)}_k(\boldsymbol{\vartheta})$. 
The likelihood is then defined as the sampling distribution of the residuals, assuming they are distributed as Gaussian noise colored by the \ac{PSD} for each detector \cite{veitch:2014wba}:
\begin{equation}
\label{eq:lik}
\mathcal{L}(d | \boldsymbol{\vartheta})\propto \exp\left[-\frac{1}{2} \sum_{k=1,2,3} \left\langle h_k(\boldsymbol{\vartheta}) - d_k \middle| h_k(\boldsymbol{\vartheta}) - d_k\right\rangle\right],
\end{equation}
where $\langle\cdot|\cdot\rangle$ denotes the noise-weighted inner product~\cite{Finn:1992wt}. Here for the Advanced LIGO 
noise spectral density we use the \texttt{ZERO\textunderscore DET\textunderscore high\textunderscore P} PSD~\cite{Shoemaker:2010}, while for Virgo we use the PSD in Ref.~\cite{2012arXiv1202.4031M}. We use the common ``zero-noise'' approximation, where instead of averaging many \ac{PDF}s 
obtained with different Gaussian noise realizations, we directly obtain this averaged \ac{PDF} by setting the noise realisation, $d_k$, to be identically zero, while keeping the detectors' PSD when computing the noise-weighted inner product in Eq.~\eqref{eq:lik}.

We follow the choices in Ref.~\cite{veitch:2014wba} for the prior probability density $p(\boldsymbol{\vartheta})$ 
in Eq.~(\ref{eq:bayes}). When recovering the signal with the pEOBNR model, we sample the QNM complex frequencies in 
the dimensionless parameter $G M_{\rm BH}\sigma_{\ell m}/c^3$ with a flat prior $GM_{\rm BH}\omega_{\ell
  m}/c^3\in [0.3,1]$ and $GM_{\rm BH}/\tau_{\ell m}/c^3\in [0.03,0.2]$, where
$M_{\rm BH}$ is the mass of the remnant BH. These priors are chosen such that
within this range, the pEOBNR model is reasonably smooth at the matching
point between the inspiral-plunge and merger-ringdown parts. When we use the
damped sinusoids, we employ flat priors for the dimensionful quantities
$f_{\ell m}\in [50,500] \rm{Hz}$ and $1/\tau_{\ell m}\in [50,500]
\rm{s}^{-1}$, with $2\pi f_{\ell m}=\omega_{\ell m}$. Finally, for all runs done, we have not seen that the posteriors for the frequency and damping time of the 220 or 330 modes lean against the prior boundaries, whenever the SNR after merger of the corresponding mode is above $\sim 5$.

%%%%%%%%%%%%%%%%%%%%%%%%%%%%%%%%%%%%%%%%%
\begin{figure}[ht]
\begin{center}
\includegraphics[width=0.48\textwidth]{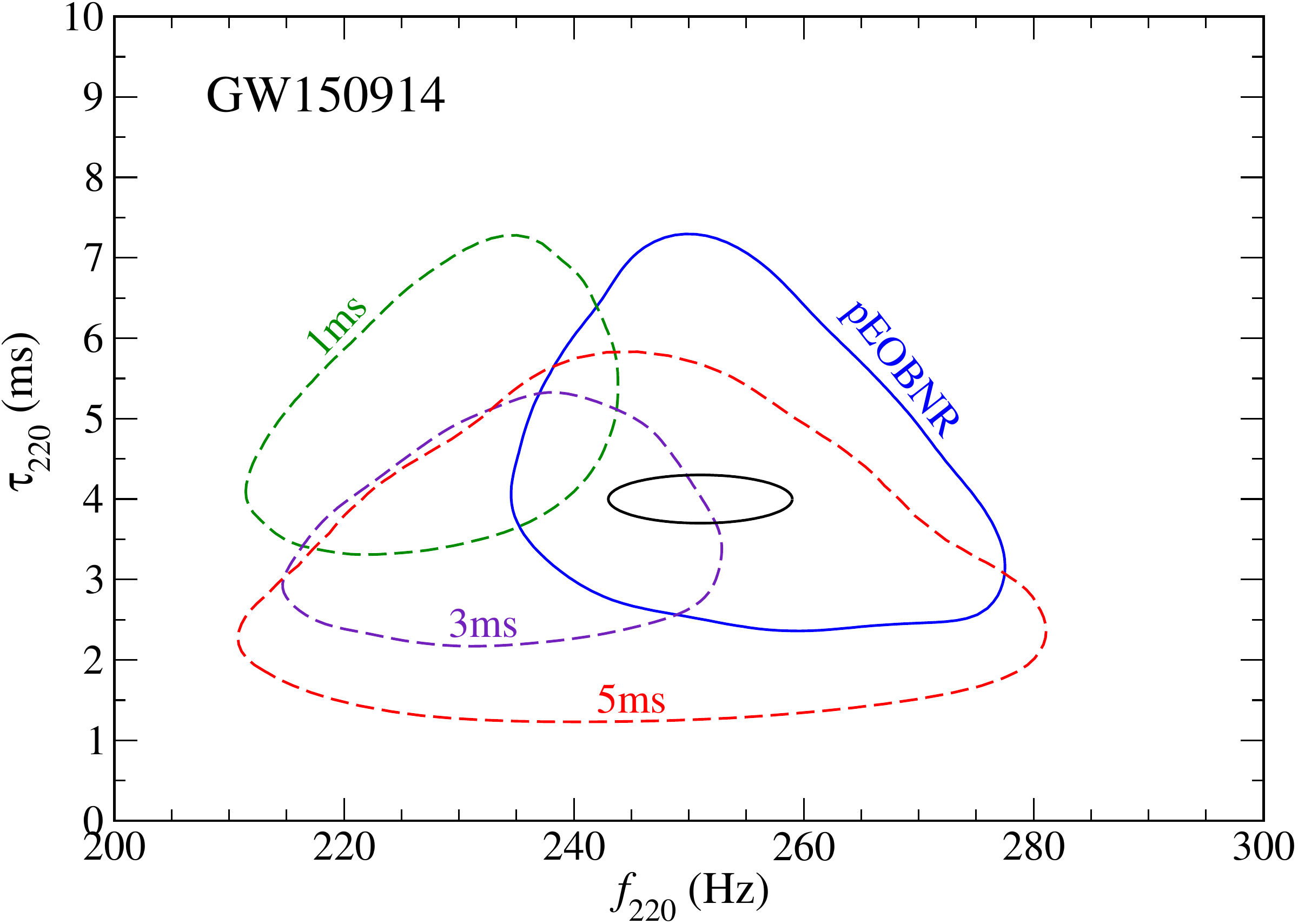}
\caption{90\% credible interval contours for the dominant QNM, using the pEOBNR model and a damped sinusoid model at starting times $t_{\rm RD} = 1, 3, 5$ ms after merger.
 The black solid line shows the 90\% credible region for the frequency and decay time of the (220) QNM inferred from the posterior distributions of the remnant BH mass and spin parameters, as derived in Ref.~\cite{TheLIGOScientific:2016src}.
GW150914 is consistent with the coalescence of two nonspinning BHs, with an inferred total (redshifted) mass of $M/M_{\odot}=70.6^{+4.6}_{-4.5}$, mass ratio $q=0.82^{+0.17}_{-0.20}$ and luminosity distance $D_L/{\rm Mpc}=410^{+160}_{-180}$~\cite{TheLIGOScientific:2016wfe}.
  \label{fig:GW150914}}
\end{center}
\end{figure}
%%%%%%%%%%%%%%%%%%%%%%%%%%%%%%%%%%%%%%%%%

%%%%%%%%%%%%%%%%%%%%%%%%%%%%%%%%%%%%%%%%%
\begin{figure*}
\begin{center}
\includegraphics[width=0.48\textwidth]{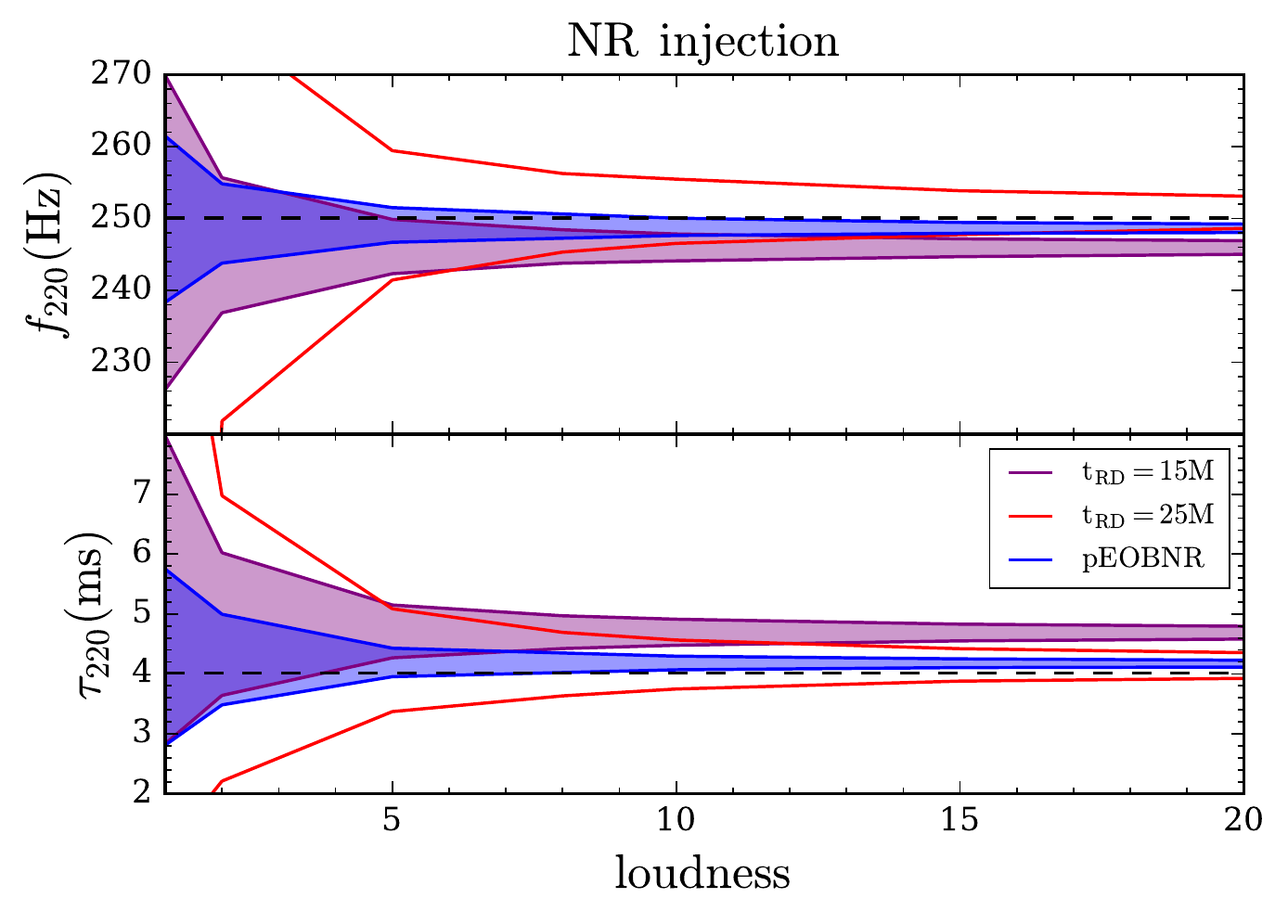}
\includegraphics[width=0.48\textwidth]{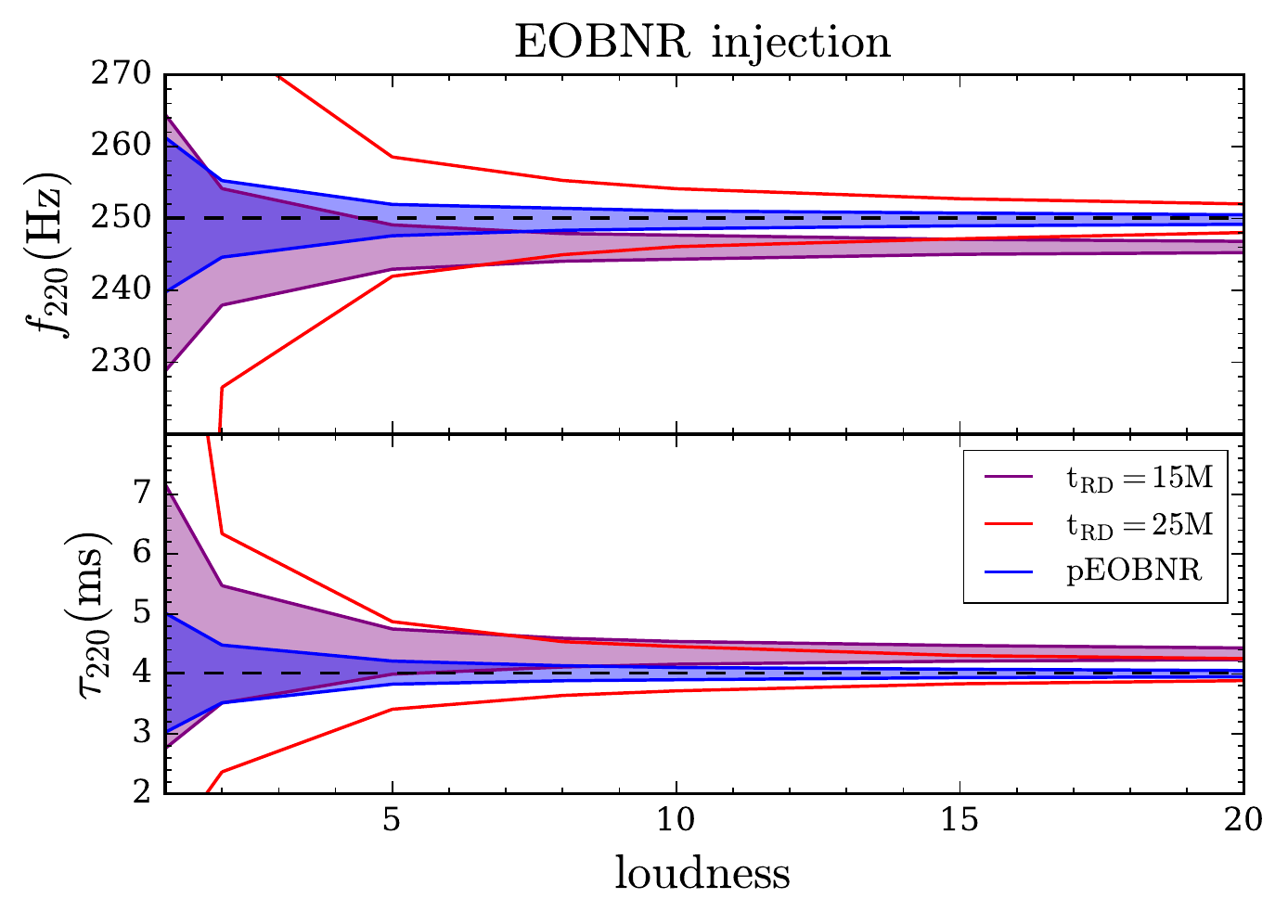}
\includegraphics[width=0.48\textwidth]{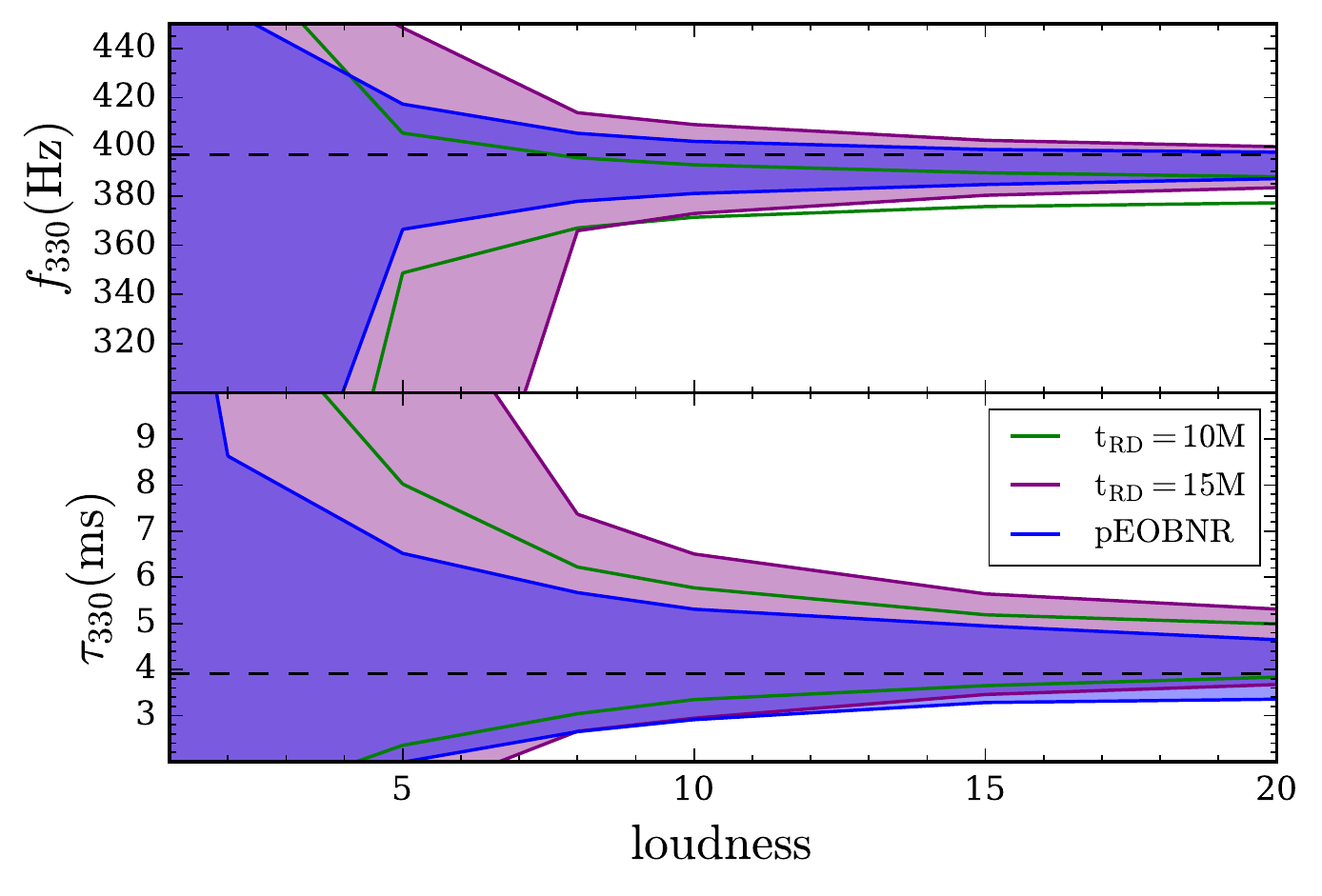}
\includegraphics[width=0.48\textwidth]{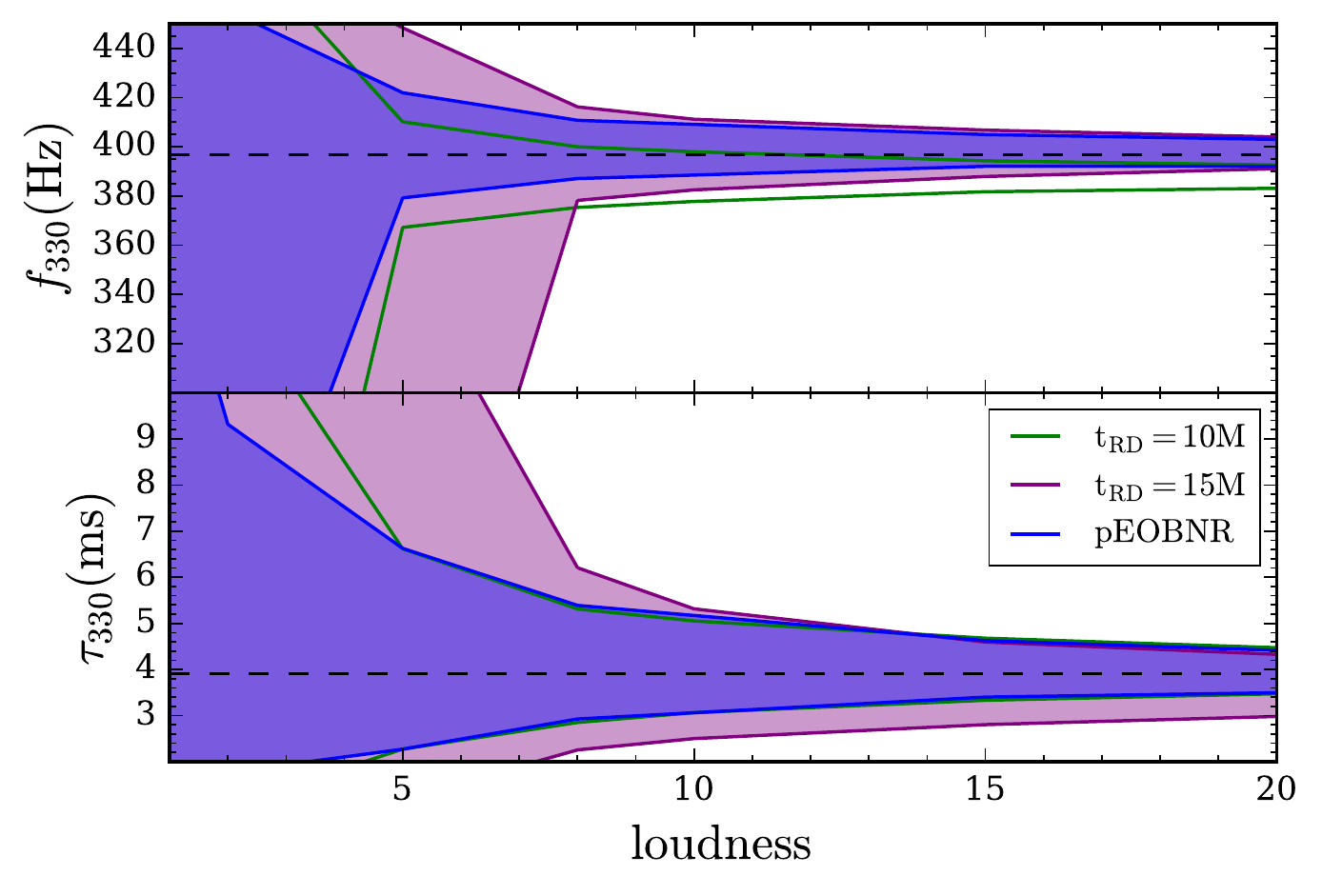} 
\caption{95\% credible interval contours for the frequency and damping time of the (220) and (330) modes of a GW event with mass ratio $q=1.5$, total (redshifted) mass $M=70M_{\odot}$ as a function of the \emph{loudness}, defined as ${\rm loudness}=500 {\rm Mpc}/D_L$. The dashed black lines corresponds to the injected values. We compare the recovery using the pEOBNR model with the one using a single damped sinusoid with starting time $t_0 = t_c + t_{\rm RD}$. In the left panels we show the recovery for an NR injection, while in the right panels we show the recovery for an injection with an EOBNR waveform with the same parameters.
  \label{fig:thrane}}
\end{center}
\end{figure*}
%%%%%%%%%%%%%%%%%%%%%%%%%%%%%%%%%%%%%%%%%

\subsection{Putting the IMR  model to test using GW150914}

GW150914~\cite{Abbott:2016blz} was the first and, so far, loudest BBH's GW
  signal detected by Advanced LIGO and Virgo. Constraints for the
  frequency and damping time of the dominant QNM for this event
  were computed in Ref.~\cite{TheLIGOScientific:2016src}. Following
  the latter, we use 8 s of data centered around GW150914 from both
  Livingston and Hanford LIGO detectors, and infer GW150914's
  parameters using the pEOBNR model. In Fig.~\ref{fig:GW150914} we
  show the 90\% credible intervals of the 2D \ac{PDF} for the recovery
  of the dominant QNM frequency $f_{220}$ and damping time
  $\tau_{220}$. We also compare the results with the constraints
  that we obtain when using the two damped sinusoid model with different starting
  times~\footnote{For comparison with Ref.~\cite{TheLIGOScientific:2016src}, we fix the starting time of the damped sinusoid
    model to be $t_0 = t_c +1; 3; 5$ ms (in units of the BBH total mass this corresponds to $\sim 3M; 9M; 15M$ after merger, respectively), where we choose $t_c$ to be
    given by the maximum likelihood GPS time obtained from the run
    using the pEOBNR model, namely we use $t_c=1126259462.408\, \rm{s}$.
    For the sky position we fix the right ascension 
    $\alpha=1.953\, \rm{rad}$ and declination $\delta=-1.2\,
    \rm{rad}$.}.
    We also show the frequencies as inferred by
  assuming GR and using the posterior distributions of the remnant
  mass and spin parameters as derived in
  Ref.~\cite{TheLIGOScientific:2016src} (black solid line). Our main
  conclusion is that the pEOBNR model gives constraints that are in full
  agreement with the ones inferred from the posterior distributions of
  the remnant mass and spin parameters, and even slightly stronger than
  the damped sinusoid model. In addition, as already emphasized, the
  pEOBNR model avoids intrinsic issues inherent with using a damped sinusoid
  model such as potential biases due a non-optimal choice of the
  \emph{a priori} unknown starting time for the ringdown signal. In particular, one can see that choosing the damped sinusoid model to start $t_{\rm RD}=1$ ms after merger gives inconsistent results with the expected frequency and damping time, showing that this choice is too early for the start of the ringdown, something which is \emph{a priori} unknown from the data alone. In
  addition, the uncertainty in the measurement of the time at coalescence and sky position is naturally included in the pEOBNR model,
  while such uncertainty cannot be easily incorporated in the damped
  sinusoid model (see Ref.~\cite{Cabero:2017avf} for a proposal on how
  to include such uncertainty).

%%%%%%%%%%%%%%%%%%%%%%%%%%%%%%%%%%%%%%%%%
\begin{figure*}
\begin{center}
\includegraphics[width=0.48\textwidth]{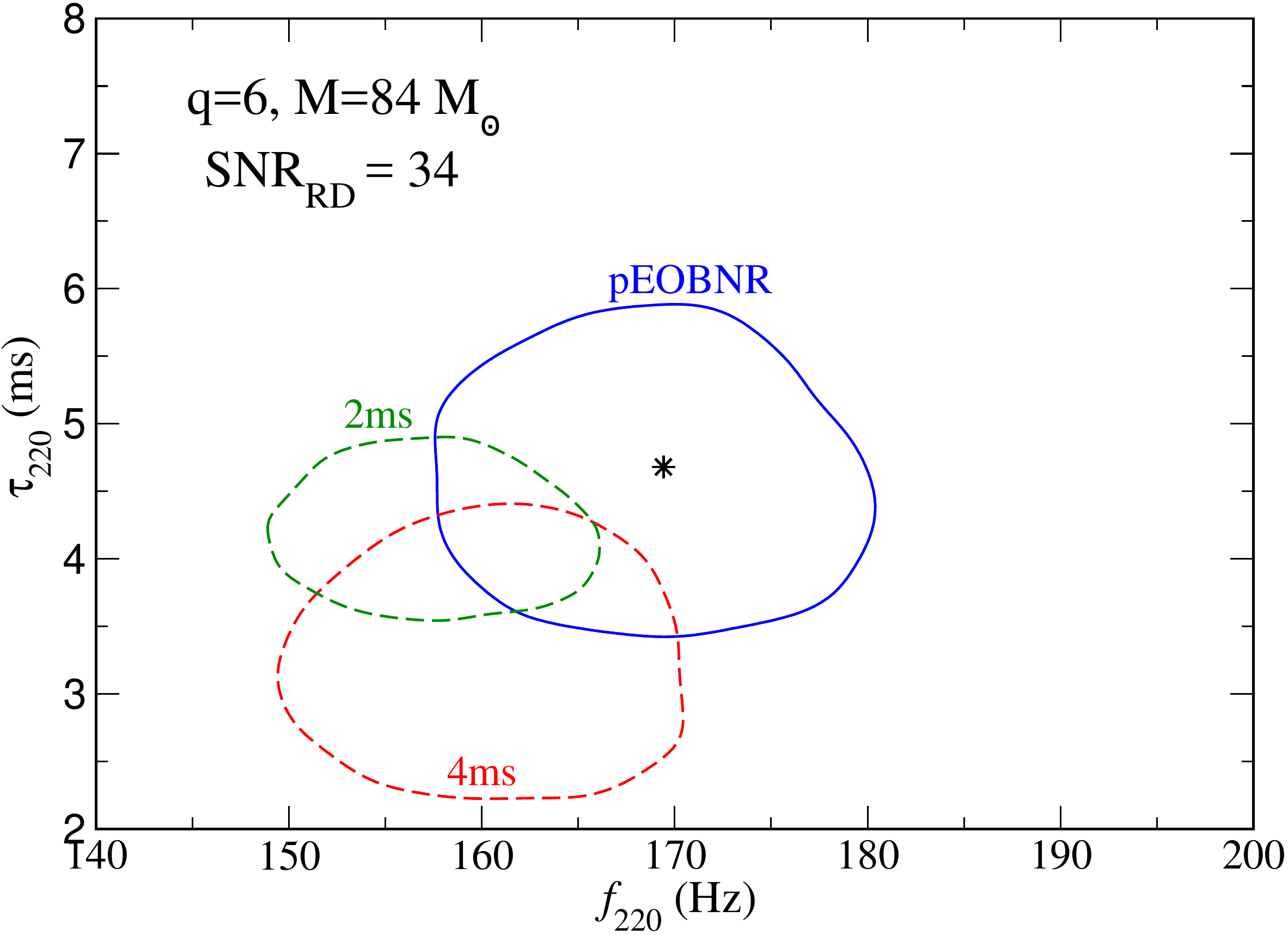}
\includegraphics[width=0.48\textwidth]{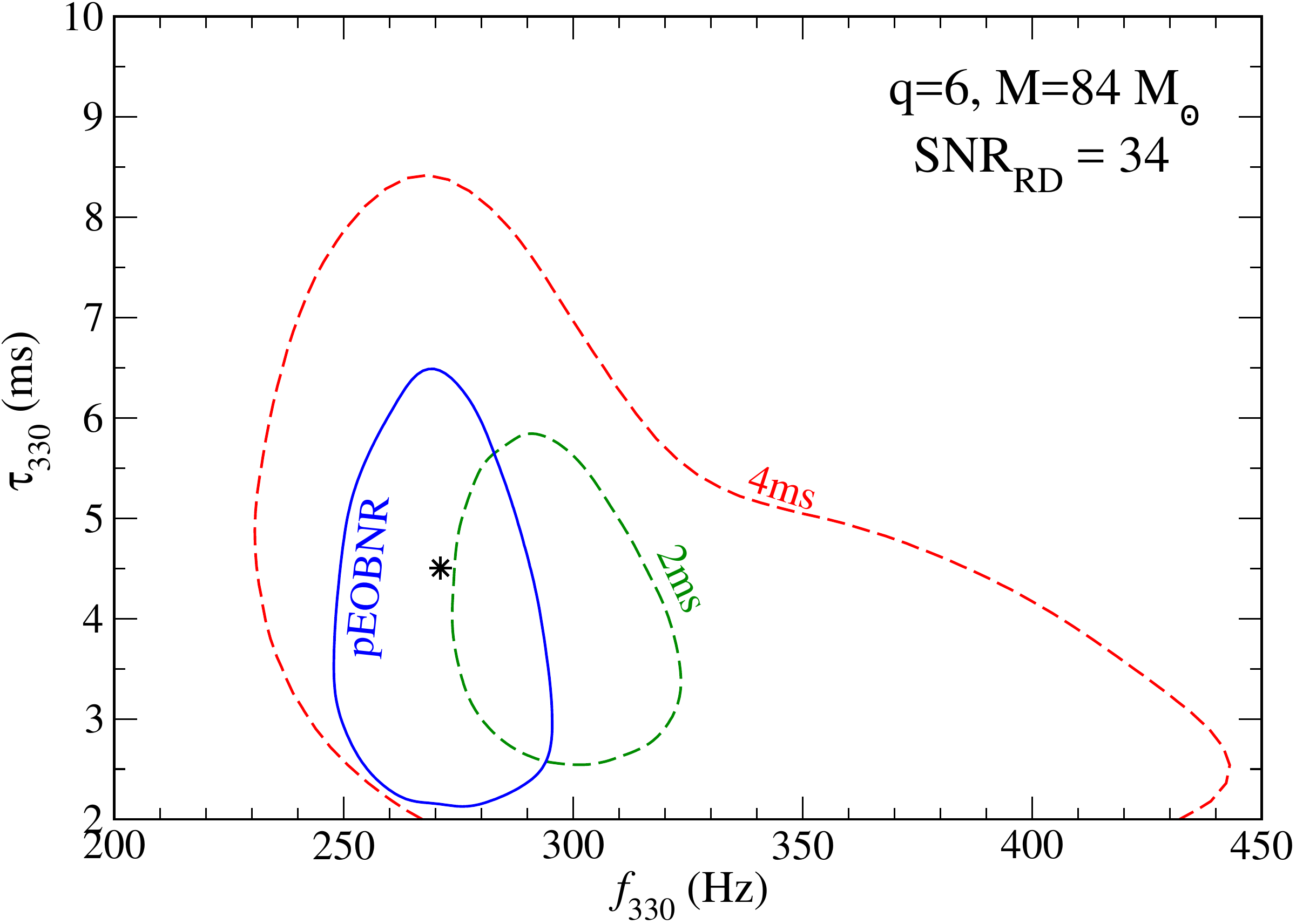}
\caption{Left: 90\% credible interval contours for the $(220)$ QNM complex frequency for NR waveform with mass ratio $q=6$, total (redshifted) mass $M=84 M_\odot$ and IMR network SNR$\approx 70$, corresponding to an SNR in the ringdown of ${\rm SNR}_{\rm RD}\approx 34$. The black star corresponds to the injected value for the $(220)$ QNM. Right: Same, but for the recovery of the $(330)$ QNM. 
  \label{fig:NR_injections_2}}
\end{center}
\end{figure*}
%%%%%%%%%%%%%%%%%%%%%%%%%%%%%%%%%%%%%%%%%

\subsection{Putting the IMR waveform model to test using numerical-relativity waveforms}

It was recently claimed in Ref.~\cite{Thrane:2017lqn} that there is an
intrinsic limit in the accuracy with which one can extract QNM
frequencies, when describing the post-merger signal by a sum of
exponentially damped sinusoids. In particular, 
Ref.~\cite{Thrane:2017lqn} argued that although a more sensitive
detector can probe later times in the GW signal, it does not necessarily
mean one can get tighter constraints on the ringdown frequencies and
damping times, due to a tension between the need to maximize the SNR
at which one extracts the QNM frequencies, and an optimal choice for
the time at which the signal can be well-described by a sum of
QNMs. The authors speculated that this effect might be due to residual
nonlinearities decaying on similar timescales to the ringdown signal,
but more recently Ref.~\cite{Baibhav:2017jhs} argued that this effect
is likely due to the increasing importance of the overtones in the
large-SNR limit.

In fact, as we show below and as expected, we do not find any conclusive evidence of this
limitation when using the IMR waveform at our disposal. In particular, as already
emphasized, the pEOBNR model includes overtones and naturally encodes
information on the starting time of the ringdown. In addition to these
features, the model also includes crucial information necessary to
accurately measure subdominant modes, such as time shifts between the
peak of the different modes and their relative phase and amplitude
difference compared to the dominant (220) mode.

To reproduce the features seen in Ref.~\cite{Thrane:2017lqn} we inject an
NR waveform with mass ratio $q=1.5$ (SXS:BBH:0007) and total
redshifted mass $M=70 M_\odot$ at different distances while keeping
all the other parameters constant\footnote{We use $\theta=2.2\,
  \rm{rad}$, $\alpha=1.21\, \rm{rad}$, $\delta=-1.165\,
  \rm{rad}$ and $t_c=1126259462$
  s.}. Following~\cite{Thrane:2017lqn} we define the loudness of the
signal as ${\rm loudness}=500 {\rm Mpc}/D_L$.  For the injections that we
consider, ${\rm loudness}=1$ corresponds to a network ${\rm
  SNR}\approx 50$ and ${\rm SNR}_{\rm RD}\approx
20$\footnote{Here we define the SNR in the ringdown, ${\rm SNR}_{\rm RD}$, 
as the SNR computed starting from the peak (or merger) of the
    $(2,2)$ mode.}. We also note that, everything else being fixed,
${\rm loudness}\propto {\rm SNR}$. Following
Ref.~\cite{Thrane:2017lqn}, and to avoid potential errors introduced by the
presence of higher-modes in the NR signal, we inject the (2,2) and
(3,3) modes of the NR waveform separately. To understand whether
potential biases are due to residual nonlinearities in the NR waveform
or simply due to a non-optimal choice of the starting time for the
damped sinusoid model, we also inject the EOBNR waveform mode~\cite{Pan:2011gk} 
with the same parameters of the NR waveform, for which the
ringdown part is exactly described by a sum of QNMs
(see Eq.~\eqref{RD}). The injected signals are then recovered using
both the pEOBNR model, which has free QNM complex frequencies, and a single
damped sinusoid model, with different starting times.

Our results are summarized in Fig.~\ref{fig:thrane}. As expected, by
increasing the loudness (i.e., increasing the SNR of the injected
signal), the error decreases roughly as $1/{\rm SNR}$. As can be seen
in the left panels, when recovering the NR signal with a single damped
sinusoid, if one chooses a starting time too early
after merger, one expects the damped sinusoid to recover inaccurate
QNM frequencies, while choosing a starting time too late after merger 
leads to large statistical errors. We find that one
needs to start the matching at a time after merger of at least $t_{\rm
  RD} \gtrsim 20M$ for the (220) mode and $t_{\rm RD} \gtrsim 15M$ for
the (330) mode, to get unbiased frequencies and damping times. This is
consistent with recent studies on the starting time of the ringdown in
BBH mergers~\cite{Bhagwat:2017tkm}. On the other hand, the pEOBNR model
recovers both the frequency and damping time of the NR waveform with a
very a good accuracy, although we find a small bias of $\sim 1\%$ for
the (220) frequency compared to the injected value. This is likely a
systematic bias due to modeling errors in the inspiral-plunge part of
the IMR model~\cite{Littenberg:2012uj}. In fact, as can be seen in the right panels, when
injecting the EOBNR waveform, as expected the pEOBNR model recovers
unbiased frequencies and damping times while the behavior of the
damped sinusoid model is similar to what we found for the NR
injection, therefore no apparent sign of residual nonlinearities in the NR waveforms are found when using the damped sinusoid model. In addition, in Ref.~\cite{Littenberg:2012uj} it was shown that at sufficiently large SNRs, biases of the same order can occur for the measured BH masses when recovering a NR waveform with an EOBNRv2HM template. We expect that this error propagates to the recovered QNM frequencies, explaining the small bias we observe for the pEOBNR model. We therefore find no conclusive evidence that the limitation discussed
in Ref.~\cite{Thrane:2017lqn} is due to residual nonlinearities in the ringdown part of the NR waveform, and in particular we find no
evidence that the IMR pEOBNR model has such limitation (aside from modeling errors).

So far, we have assumed that the different modes in the signal can be
distinguished and recovered separately. In a realistic scenario one would 
prefer instead to use the IMR pEOBNR model against the full GW signal,
since disentangling the different modes is a very challenging task
that would induce unavoidable systematic errors. Therefore, in
Fig.~\ref{fig:NR_injections_2} we also show an example where we
inject an NR waveform with all available modes (i.e., up to $\ell=8)$, for mass
ratio $q=6$ (SXS:BBH:0166) and total (redshifted) mass $M=84 M_{\odot}$. We
consider an injection with total network ${\rm SNR}\approx 70$,
corresponding to a luminosity distance $D_L=160 {\rm Mpc}$ and ${\rm
  SNR}_{\rm RD}\approx 34$.  We recover again the GW signal using the pEOBNR 
model, with all the modes available in the model, and contrast it with
the recovery when using a two damped-sinusoid model, with amplitudes
fitted to NR\cite{Gossan:2011ha}, using different starting times. Due to the large-mass
ratio, in this case there is a clear hierarchy between the amplitude of different
modes, and higher modes have a non-negligible contribution to the
overall waveform. For this mass ratio the peak amplitude of the
(3,3)-mode is roughly $70\%$ smaller than the (2,2)-mode, as can be
seen in Fig.~\ref{fig:EOBvsNR}, and therefore strong constraints on a
second QNM can be obtained even for a reasonable SNR in the ringdown (i.e., $\approx 34$). 
As we see, the pEOBNR model recovers unbiased results for the ringdown frequency and
damping time, even if the NR waveform includes more subdominant modes. On the
other hand, when using the two damped-sinusoid model and choosing starting times that give comparable errors to the
pEOBNR model, we always recover slightly biased QNM parameters. These results demonstrate 
the need of including more physical effects, e.g. include more modes and overtones~\cite{London:2014cma,London:2018gaq,Cotesta:2018fcv}, in the more theory-agnostic damped-sinusoid model, if one wanted to use it to get accurate and precise values for
the QNM frequencies and damping times of BBHs event, and test the
no-hair conjecture.

We note that at the time of this writing, no suitable NR waveform computed in alternative theories of gravity are available for testing. While the tests in this section validate our approach to constrain small deviations from GR, we do hope that further tests with non-GR waveforms will be performed in the future.

\section{Testing the general relativistic no-hair conjecture}\label{sec:no_hair}

Having laid down the ability of the pEOBNR waveform model to measure
the ringdown complex frequencies, we now investigate the capacity 
of the IMR model to detect small deviations from GR in
the ringdown part of the signal using two approaches: (i) a Bayesian
model selection scheme, and (ii) by directly measuring the QNM frequencies using Bayesian parameter estimation
and computing the constraints on deviations from GR. 

Such approaches have been used in the past~\cite{Gossan:2011ha,Meidam:2014jpa}, however focusing on the
damped sinusoid model, which as we have argued above, is prone to technical
difficulties. Therefore, from now on, we focus solely on the IMR pEOBNR model.

\subsection{Bayesian model selection}\label{sec:modelselec}

Bayesian model selection has been extensively used in the context of
testing GR~\cite{DelPozzo:2011pg,Li:2011cg,Gossan:2011ha,Meidam:2014jpa}, and
is particularly useful to find statistical evidence for deviations
from GR even when the majority of the GW events have a small SNR, and 
parameter estimation alone might not be enough to confidently
measure such deviations. Model selection can also naturally be used
to get statistical evidence from a small deviation from GR by
combining the information from several
observations~\cite{Li:2011cg,Meidam:2014jpa}. In fact, for most of the
BBH events that Advanced LIGO and Virgo is detecting, we do not expect to
be able to impose strong constraints on the QNM complex
frequencies~\cite{Berti:2016lat}, and therefore this is the most
promising avenue to detect deviations from GR, before LISA or third-generation 
detectors on the ground, such as Cosmic Explorer and Einstein Telescope 
are online. 

As said above, similar studies were done in the
past in Refs.~\cite{Gossan:2011ha,Meidam:2014jpa}, but they focused on
damped-sinusoid models, both for the injected GW signal and the waveform model 
used to recover it, and they were done using the PSD of Einstein
Telescope. Besides the use of an IMR model to recover the signal,
another crucial difference here, is that we also inject IMR waveforms. 
If one would do a Bayesian model selection study
on such population using damped sinusoids as templates, one would need
to deal with the problem of defining the optimal starting time for the
ringdown, that is in general dependent on the particular binary's  
configuration. Using the IMR model completely avoids this problem. In
addition, a Bayesian model selection with an IMR model also
naturally incorporates the consistency test that both the inspiral-plunge and
merger-ringdown are consistent with GR. 

In general, given some observed data $d$, the support for a given model hypotheses $\mathcal{H}$ can be quantified by integrating Eq.~\eqref{eq:bayes} (with $h$ replaced by $\mathcal{H}$) over $\boldsymbol{\vartheta}$:
\begin{equation}
p(\mathcal{H}|d) \propto  \mathcal{L}(d|\mathcal{H}) \times p(\mathcal{H}).
\label{eq:bayesMS}
\end{equation}
To compare two different model hypotheses, say $\mathcal{H}_i$ and $\mathcal{H}_j$,  in light of the observed data, we compute the ratio of posterior probabilities also known as the odds ratio~\cite{DelPozzo:2011pg,Li:2011cg}:
\begin{equation}
\mathcal{O}_{j}^{i} = 
\frac{p(\mathcal{H}_{i} |d)}{p(\mathcal{H}_{j} |d)}=
\frac{p(\mathcal{H}_{i})}{p(\mathcal{H}_{j})}\frac{\mathcal{L}(d|\mathcal{H}_{i})}{\mathcal{L}(d|\mathcal{H}_{j})}=
\frac{p(\mathcal{H}_{i})}{p(\mathcal{H}_{j})}B_{j}^{i}\,,
\end{equation}
where $p(\mathcal{H}_{i})/p(\mathcal{H}_{j})$ is the prior odds of the two hypotheses and $B_{j}^{i}$ is the Bayes factor.
%In the following we use  $p(\mathcal{H}_{i})/p(\mathcal{H}_{j})=1$.
%By construction, if $\mathcal{O}_{j}^{i}>1(<1)$ the data prefers the model $i (j)$.}
In the following, we quote directly the Bayes factor, so that by construction, if $B_{j}^{i}>1(<1)$ the data prefers the model $i (j)$. Then, we need to multiply by the prior odds (which in the case of GR versus non-GR could be a large effect) to get the odds ratio.

Even though no waveform model that corresponds to a non-GR theory is currently available, we may ask: ``Given the observed data, are the QNM frequencies and damping times compatible with GR?''. To address this question, we 
consider two different hypotheses models: (i) $\mathcal{H}_{\rm GR}$, which corresponds to the 
hypothesis that the events are described by EOBNR waveforms where QNM frequencies are fixed to the 
GR values, and (ii) $\mathcal{H}_{\rm nonGR}$, which corresponds to the
hypothesis that the QNM complex frequencies are (additional) free parameters and 
are described by pEOBNR waveforms. Note that the latter also includes GR for a particular choice of QNM frequencies, however, even if GR is the correct theory, the model is penalized when performing Bayesian model selection due to the addition of extra parameters that are not needed to describe the data. 
For simplicity, in this work, 
the model $\mathcal{H}_{\rm nonGR}$ uses the hypothesis that only 
the frequencies and damping times of the (220) and (330) are not fixed by 
the inspiral parameters as given in GR, but all the other QNMs included in the model 
do (i.e., the 21-mode, 44-mode, and 55-mode and their overtones).
We note that we could follow an approach similar to TIGER
(Test Infrastructure for GEneral Relativity)~\cite{Li:2011cg}, where
all combinations of possible free parameters are included in the
non-GR hypothesis. This approach is in general quite robust in finding
deviations from GR even for low SNR systems, but it can be computationally
expensive because several models must be analyzed. Therefore, for practical purposes,  we only consider the hypothesis that the frequencies and damping times of the (220) and (330) are free at the same time.

To carry out the analysis on 
a reasonable timescale, we 
fix the sky position and the parameters influencing mostly the inspiral-plunge phase, namely the mass ratio
$q$ and chirp mass $M_c$. Given that the inspiral is the same for both
the GR (EOBNR) and non-GR (pEOBNR) hypotheses, this is a reasonable assumption that
should not influence the qualitative picture of the results, especially 
at large SNRs, where the inspiral parameters and the sky position are
measured with very good accuracy. However, the model and framework presented here 
are not limited to those assumptions, and we plan to relax them and 
do a more comprehensive analysis in the near future.

Given a detection, we compute the Bayes factor as:
\be
B^{\rm nonGR}_{\rm GR}=\frac{B^{\rm nonGR}_{\rm noise}}{B^{\rm GR}_{\rm noise}}\,,
\ee
where $B^{\rm nonGR}_{\rm noise}$ and $B^{\rm GR}_{\rm noise}$ are the Bayes factors for $\mathcal{H}_{\rm nonGR}$ and $\mathcal{H}_{\rm GR}$ against the hypothesis that
the data contain only noise, which we obtain using a nested sampling algorithm as implemented in the \textsc{LIGO Algorithm Library}~\cite{veitch:2014wba}. 

For the catalogs of injections we construct two populations of 100
BBH sources, one with GR waveforms using the EOBNR
waveform model~\cite{Pan:2011gk}, that we call the GR population,
and a second catalogue with the pEOBNR model with QNM 
frequencies given by $\sigma_{220}=\sigma^{\rm
  GR}_{220}(1+\delta\sigma)$ and $\sigma_{330}=\sigma^{\rm
  GR}_{330}(1+\delta\sigma)$ where we fixed $\delta\sigma=0.1$ (the same choice was done in Ref.~\cite{Meidam:2014jpa}). Below
we refer to the latter as the non-GR population. We note that this choice is not necessarily unrealistic, since deviations of $10\%$ in the QNM frequencies have been found in some alternative theories to GR. QNM frequencies of spherically
symmetric solutions were computed in theories such as
Einstein-Maxwell-dilaton~\cite{Ferrari:2000ep}, dynamical
Chern-Simons gravity~\cite{Molina:2010fb}, Einstein-dilaton-Gauss-Bonnet
gravity~\cite{Pani:2009wy,Blazquez-Salcedo:2016enn,Blazquez-Salcedo:2017txk}
and for some solutions in massive
(bi)gravity~\cite{Brito:2013wya,Brito:2013yxa,Babichev:2015zub}. On
the other hand, not much progress has been made to compute QNMs for
spinning BHs in alternative theories to GR, the only exception
being the Kerr-Newman case in Einstein-Maxwell
theory~\cite{Pani:2013ija,Pani:2013wsa,Mark:2014aja,Dias:2015wqa}. Most
of the estimates for QNMs of spinning BHs in modified gravity have
instead used the connection between the light ring and
QNMs~\cite{Blazquez-Salcedo:2016enn,Glampedakis:2017dvb,Jai-akson:2017ldo,Glampedakis:2017cgd},
which is formally only valid in the eikonal $\ell \to \infty$ limit
and known to fail to describe some families of QNMs when additional
degrees of freedom are present~\cite{Blazquez-Salcedo:2016enn}.

We draw the component (redshifted) masses of the 100 sources from a uniform
distribution between 30 and 180 $M_{\odot}$ and maximum total (redshifted) mass
$210 M_{\odot}$. This choice implies a distribution for the mass ratios proportional to $1/q^2$ with a maximum value $q=6$. We draw the sky position and orientations
$(\alpha,\delta,\psi,\theta)$ from uniform distributions on
the sphere. The signals are distributed uniformly in volume with a
network SNR for the IMR signal ranging from ${\rm SNR}=8$ to
${\rm SNR}=100$ (corresponding to luminosity distances from roughly
$D_L=100\,{\rm Mpc}$ up to $D_L=5000\,{\rm Mpc}$).

\begin{figure}[t]
\begin{center}
\includegraphics[width=0.48\textwidth]{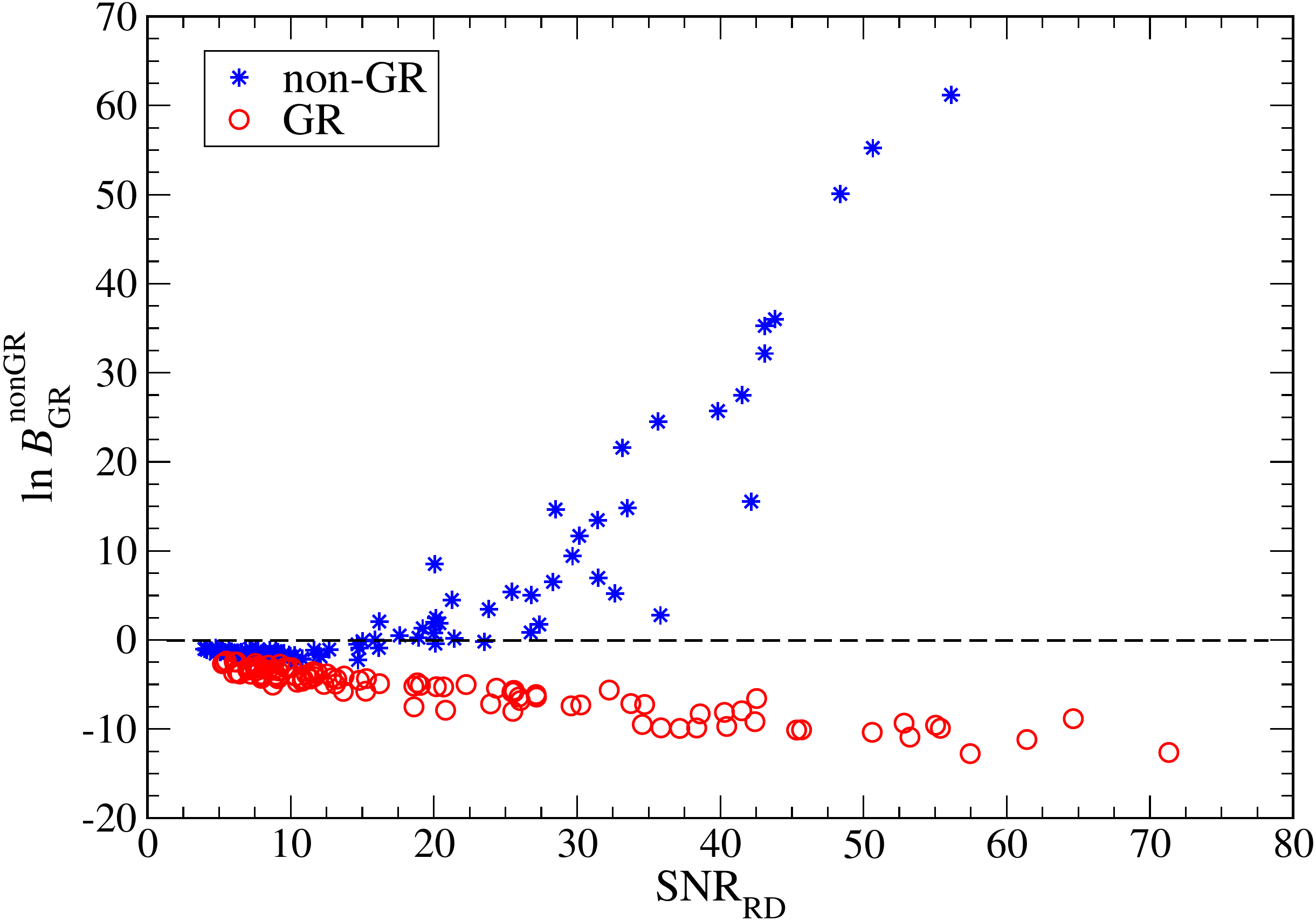}
\caption{The log Bayes factors for individual sources. The red circles represent signals with GR waveforms (EOBNR), while the blue crosses correspond to the non-GR waveforms (pEOBNR). A separation between the two is visible for ${\rm SNR}_{\rm RD}\sim 15$, and becomes more pronounced as the SNR increases.
  \label{model_selection}}
\end{center}
\end{figure}

We summarize the results in Fig.~\ref{model_selection} where we show the
(log) Bayes factor for the individual sources as a function of the SNR
in the ringdown part of the signal only (SNR$_{\rm RD}$). Since the sources are
distributed uniformly in volume, the majority of our signals has an
${\rm SNR}_{\rm RD}<10$. In this region, there is no clear difference
between the log Bayes factor for the GR and non-GR population. In fact,
for ${\rm SNR}_{\rm RD}<10$, even for the non-GR population the
preferred model is the GR waveform (which follows from the fact that
$\ln B^{\rm nonGR}_{\rm GR}<0$ for the non-GR
population). This is consistent with the fact that the GR and non-GR
waveforms have the same inspiral. Therefore, since the SNR in the
ringdown is small, and Bayesian model selection naturally incorporates
an Occam's razor selection, the model with fewer parameters (i.e., the
GR waveform) is favored in this region. However, for ${\rm SNR}_{\rm
  RD}\gtrsim 15$ we see a separation between the GR injections and the
non-GR injections and for ${\rm SNR}_{\rm RD}\gtrsim 25$, the non-GR
waveform are always favored for the non-GR events (i.e., $\ln
B^{\rm nonGR}_{\rm GR}>0$).  As one would expect, the
separation becomes much clearer with increasing ${\rm SNR}_{\rm RD}$.
We note that the threshold ${\rm SNR}_{\rm RD}$ at which deviations
from GR can be detected are dependent on the particular non-GR
deviation. However, this study illustrates the nontrivial fact that
even at relatively low SNRs, Bayesian model selection is able to find
statistical evidence for deviations from GR.

\subsection{Bounding free parameters of the ringdown signal} 

Given a set of detected GW signals from BBHs for which QNM frequencies
and damping times can be measured, the natural steps to follow are to
first test the compatibility of the waveform with GR using Bayesian
model selection, as done in the previous subsection, and then quantify
how well we can constrain deviations from GR using parameter
estimation. This can be done for single GW events, but stronger
constraints can be obtained by combining the information from all the
detections as shown in Ref.~\cite{Meidam:2014jpa}. There, two
different approaches were proposed: (i) the odds ratio obtained in the
previous subsection can be combined by just multiplying the odds ratio
coming from all the events, thus allowing to get stronger evidence for
or against GR. For a large group of $\mathcal{N}$ identical events,
this method effectively improves the SNR of the single event case by a
factor $\sim \mathcal{N}^{1/4}$~\cite{Yang:2017zxs}; and (ii) assuming
that the Bayesian model selection test gives no evidence for
deviations from GR, one combines the posterior density
distributions for $\delta\sigma_{\ell m}$, which measures the
fractional deviation from the QNM complex frequencies of a Kerr BH in
GR:
\be\label{error}
\sigma_{lm}=\sigma^{\rm GR}_{lm}(1+ \delta \sigma_{lm})\,.
\ee
Given that in GR $\delta \sigma_{lm}=0$, the information from multiple events can be combined by multiplying the posterior density distributions of all detections as
\begin{equation}\label{stacking}
p( \delta \sigma |\mathcal{H}, d_1,d_2,d_3, \ldots,\mathcal{N})=\frac{1}{p(\delta \sigma )^{1-\mathcal{N}}}\prod_{A=1}^{\mathcal{N}}p(\delta \sigma |\mathcal{H}, d_A)\,, 
\end{equation}
where $\mathcal{N}$ denotes the number of detections. For a large
  group of $\mathcal{N}$ identical events, the width of this \ac{PDF} decreases
  as $\sim\mathcal{N}^{-1/2}$. We emphasize that when using Eq.~\eqref{stacking} one assumes that the value of $\delta \sigma_{lm}$ is the same across all events. 
  Therefore, since for generic theories of gravity the deviations $\delta \sigma_{\ell m}$ could also be a function of the final BH
  mass, spin and any other charges that may be present in the {\it correct} theory of gravity, constraints obtained using
  this method only make sense if no evidence for deviations from GR are found after performing the Bayesian model selection
  test~\cite{Meidam:2014jpa}.

More recently Ref.~\cite{Yang:2017zxs} proposed an alternative
  hypothesis testing method that makes use of the combined information
  from multiple detections and could, in principle, enhance the
  efficiency to detect sub-leading modes compared to the Bayesian
  model selection method used in Ref.~\cite{Meidam:2014jpa}. This
  method proposes to make full use of the information coming from the
  measured BBH parameters, to coherently sum the ringdown signal of a
  target mode from multiple events. It could, in an ideal scenario,
  effectively improve the SNR of a single event by a factor
  $\sim\mathcal{N}^{1/2}$, assuming $\mathcal{N}$ identical
  events~\cite{Yang:2017zxs}. However, implementing the coherent stacking
  method of Ref.~\cite{Yang:2017zxs} is technically very
  challenging. Here, we follow Ref.~\cite{Meidam:2014jpa}
  and use Eq.~\eqref{stacking} to combine the information from a
  population of detected BBHs.

\begin{figure}[t]
\includegraphics[width=0.48\textwidth]{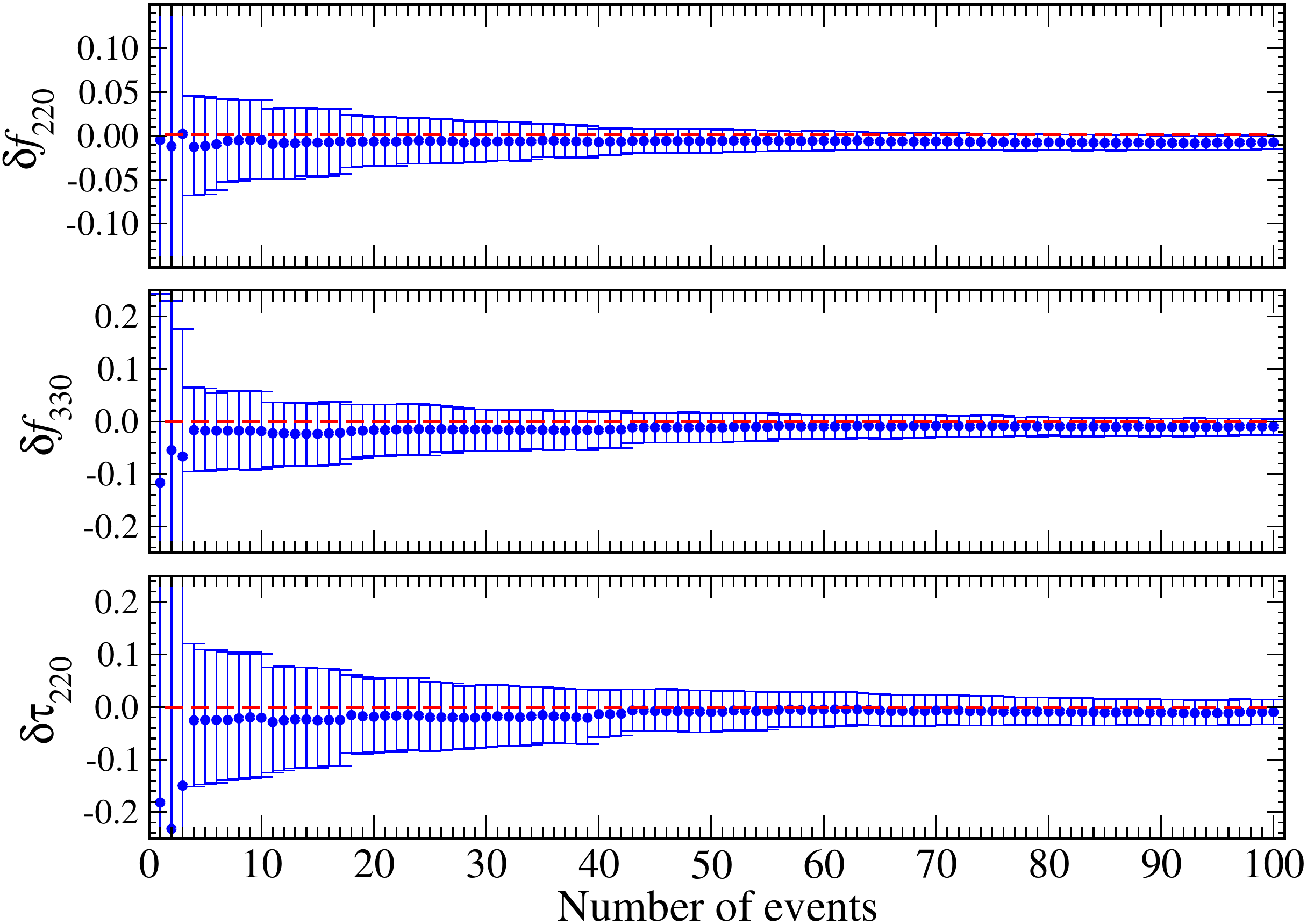}
\caption{Evolution of the medians and $95\%$ confidence intervals for deviations on the QNM frequencies as a function of the number of events included in the computation of the joint posterior density distributions, for the population of GR BBHs described in the subsection~\ref{sec:modelselec}. 
\label{fig:stacking}}
\end{figure}

\begin{figure}[t]
\includegraphics[width=0.48\textwidth]{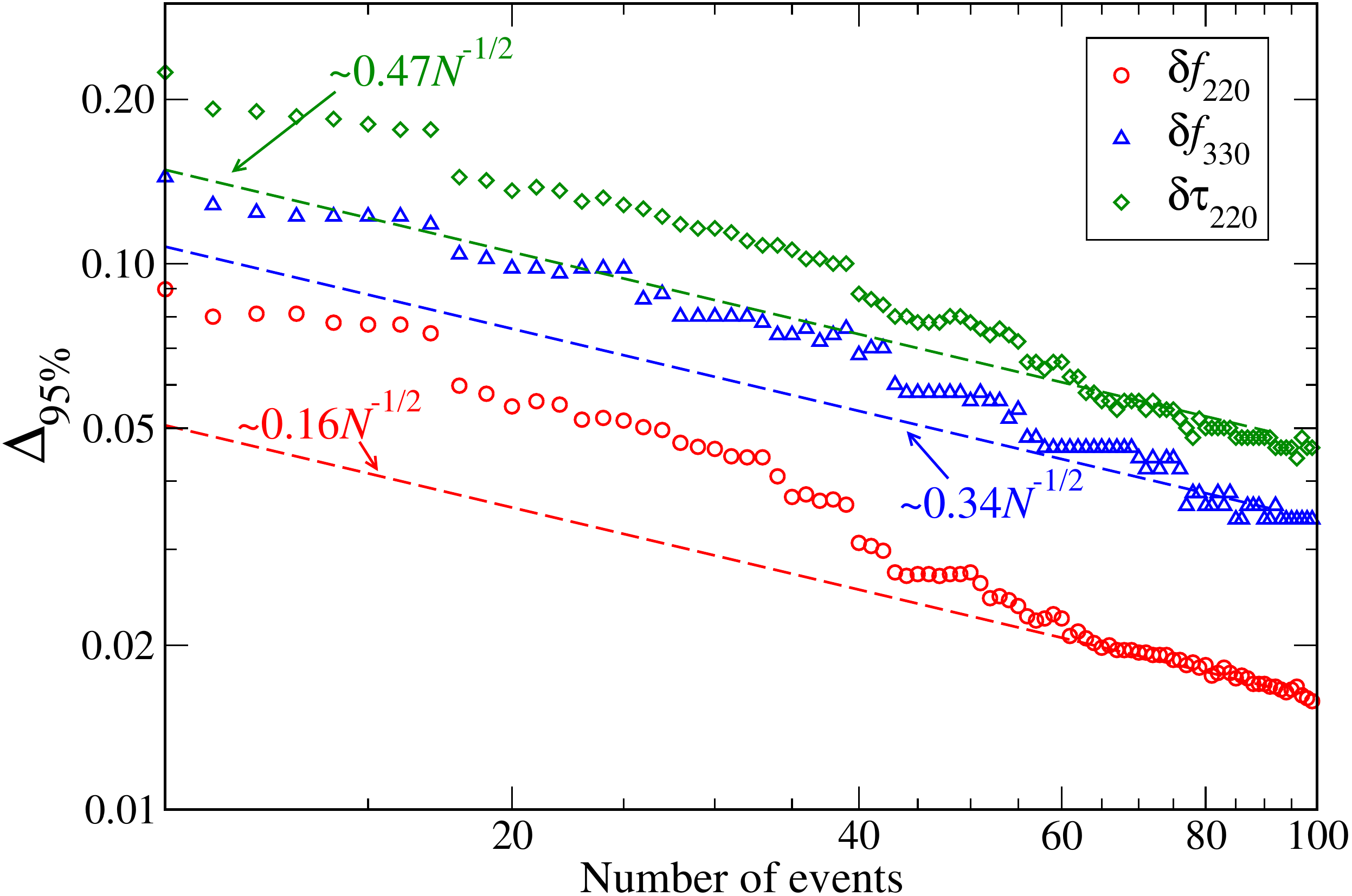}
\caption{Evolution of the width of the $95\%$ confidence intervals ($\Delta_{95\%}$) shown in Fig.~\ref{fig:stacking} as a function of the number of events $\mathcal{N}$. Note that a log-log scale is used in this plot to make apparent the power-law scaling of the error at large $\mathcal{N}$. The dashed lines scale as $\sim \mathcal{N}^{-1/2}$ showing that at large enough $\mathcal{N}$, the error decreases approximately as $\mathcal{N}^{-1/2}$. 
\label{fig:stacking_scaling}}
\end{figure}

Since for each event we sample on the parameter $\sigma_{lm}$, we compute the \ac{PDF}s for $\delta \sigma_{lm}$ a posteriori by using Eq.~\eqref{error}. To compute $\sigma^{\rm GR}_{lm}$ we use the fitting formulas in Ref.~\cite{Berti:2005ys} (see Appendix E therein) where for the spin and mass of the final BH we 
employ the fitting formulas in Ref.~\cite{Pan:2011gk} [see Eqs. (29a) and (29b) therein].  The results for the constraints on the parameters $\delta \sigma_{\ell m}$, when considering the GR BBH population described in the previous subsection\footnote{We note that for this study, unlike what was done in the previous subsection, we keep all waveform's parameters free.}, are displayed in Fig.~\ref{fig:stacking}. In particular, 
we show how the median and $95\%$ confidence interval evolve with the number of detections
ordered randomly. Although the constraints from a single event can be
quite uninformative, when all sources are taken into account the
$95\%$ confidence interval shrinks to a maximum error away from the
median of $\sim 0.7\%$, $\sim 1.6\%$ and $\sim 2.4\%$ , for $\delta 
f_{220}$, $\delta f_{330}$ and $\delta \tau_{220}$, respectively. As expected and as shown in Fig.~\ref{fig:stacking_scaling}, we find that 
at large enough $\mathcal{N}$, the error decreases approximately as $\mathcal{N}^{-1/2}$. Overall, our results are 
consistent with previous studies~\cite{Meidam:2014jpa}, although we remind that Ref.~\cite{Meidam:2014jpa} used damped sinusoids for both the injected GW signal and the recovery, while we injected and recovered with an IMR waveform that consistently includes time and phase shifts between QNMs.

It is worth noticing that if we consider only events with (total) SNR below 
30 (which accounts for 60 events of the entire population), and combine them, we 
obtain at $95\%$ confidence that the maximum errors away from the median are 
$\sim 1.7\%$ $\sim 5.3\%$ and $\sim 6.7\%$, for $\delta f_{220}$, $\delta f_{330}$ 
and $\delta \tau_{220}$, respectively. Moreover, we find that $\delta f_{220}$ is the quantity for which we 
gain less by combining several events, because it is the best measured quantity --- e.g., 
for some individual events with SNR less than 30, we get errors on the order of $\sim 5\%$. By contrast, if we 
consider only events with SNR less than 30, the errors of $\delta f_{330}$ and $\delta \tau_{220}$ 
for individual events are always larger than $20\%$.
\begin{figure}[htb]
\includegraphics[width=0.48\textwidth]{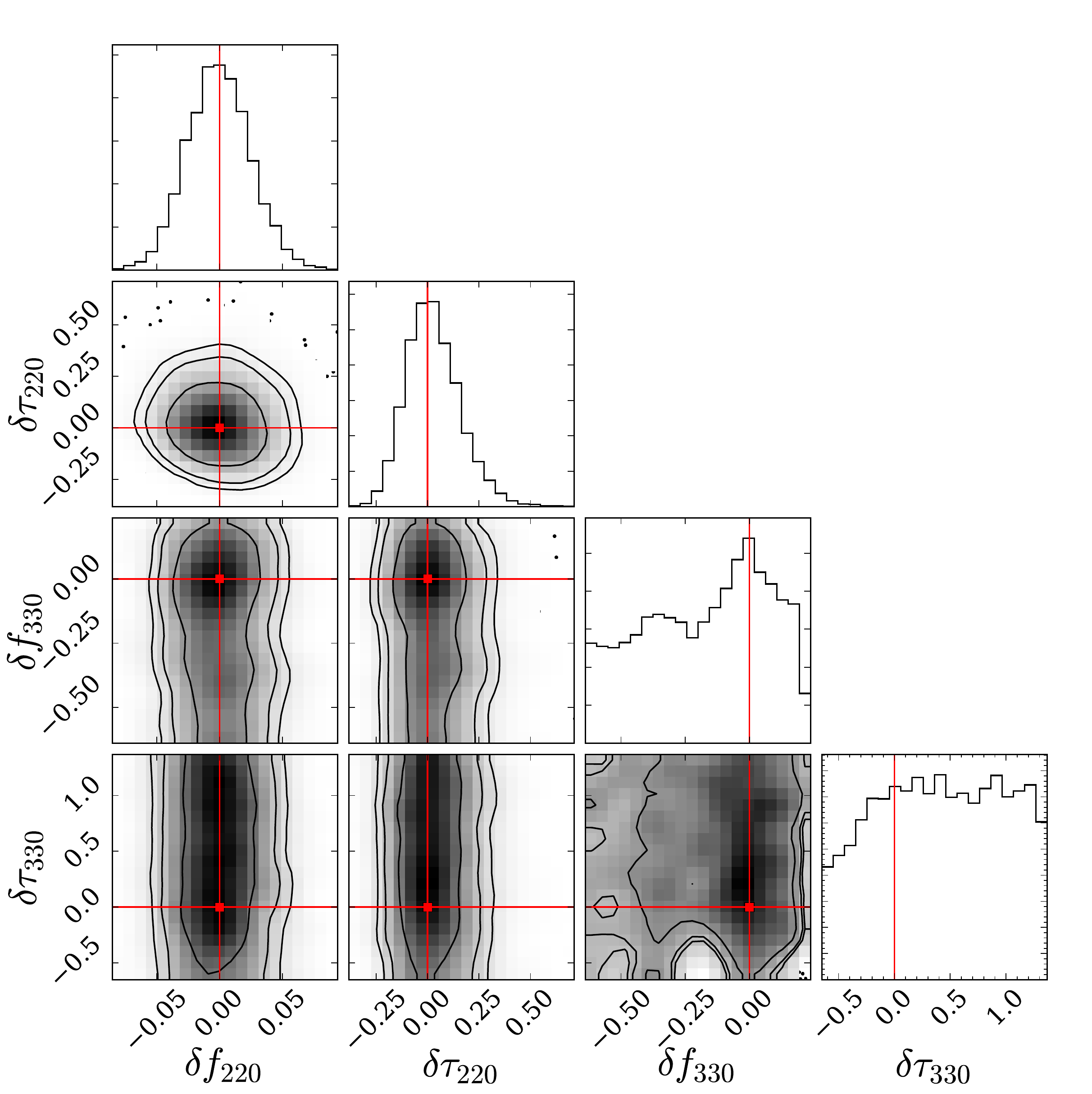}
\caption{Posterior density distributions of the quantities $\delta f_{220}$, $\delta f_{330}$, $\delta \tau_{220}$ and $\delta \tau_{330}$ for a single GW150914-like event with mass ratio $q=1.5$, total (redshifted) mass $M=70 M_{\odot}$, luminosity distance $D_L=500 {\rm Mpc}$ and inclination $\theta=2.2\,{\rm rad}$. The red solid lines correspond to the injected values.
\label{fig:corner_plot}}
\end{figure}

\begin{figure}[htb]
\includegraphics[width=0.48\textwidth]{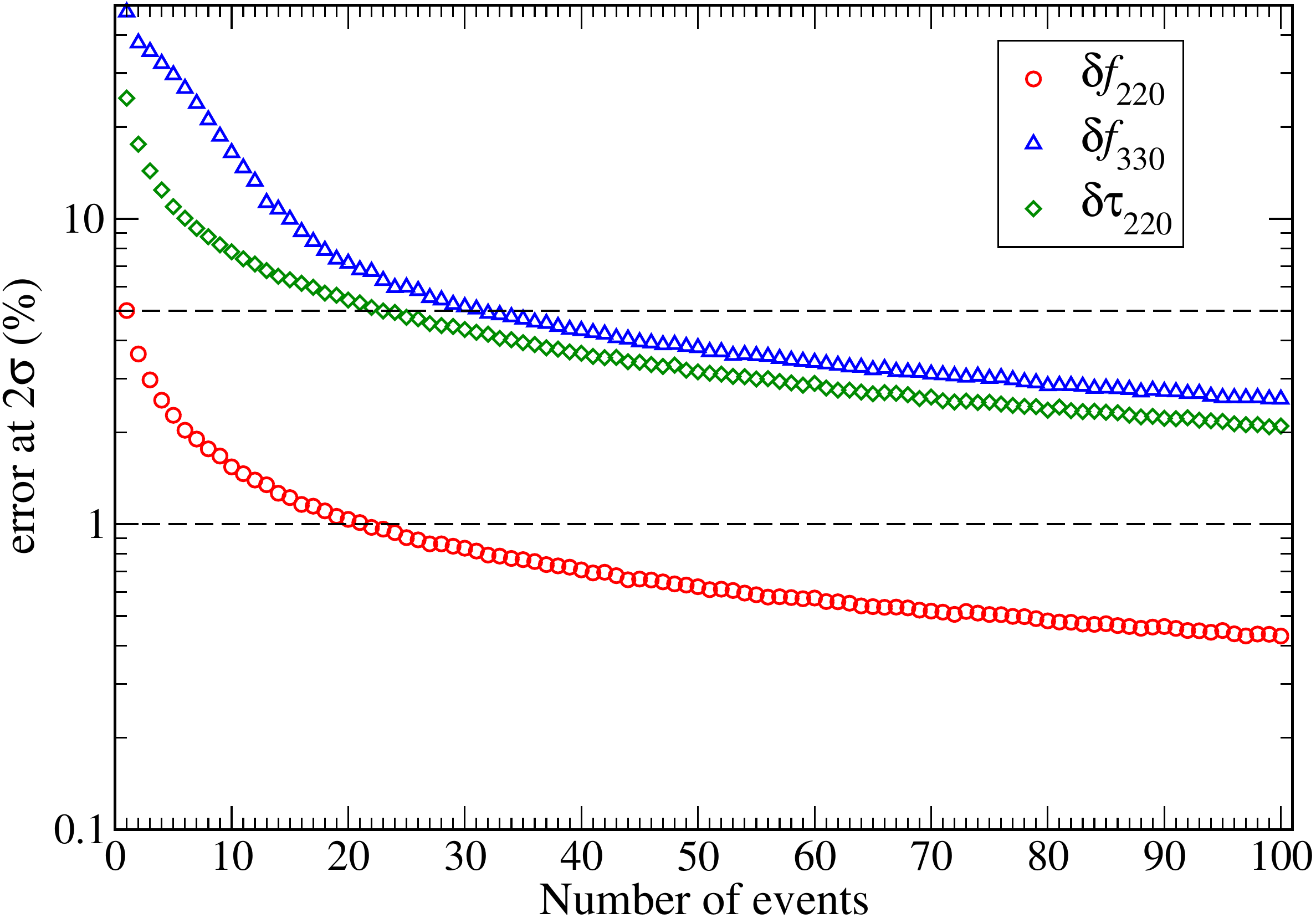}
\caption{Evolution of  2-$\sigma$ error of the joint posterior density distributions for the quantities $\delta f_{220}$, $\delta f_{330}$, $\delta \tau_{220}$, assuming identical GW150914-like events with posterior density distributions for a single event given in Fig.~\ref{fig:corner_plot}. The dashed black lines correspond to errors of $5\%$ and $1\%$.
\label{fig:GW150914_stacking}}
\end{figure}

Quite interestingly, using Eq.~\eqref{stacking} for identical GW150914-like events with
mass ratio $q=1.5$, total (redshifted) mass $M=70 M_{\odot}$, luminosity distance
$D_L=500 {\rm Mpc}$ and inclination $\theta=2.2\,{\rm rad}$
(i.e., the EOBNR injection with loudness$=1$ in Fig.~\ref{fig:thrane}), one can estimate how many such events would be needed to test the BH's no-hair conjecture with Advanced LIGO and Virgo at design sensitivity, assuming that GR is the correct theory. The posterior density distributions for a single event is shown in Fig.~\ref{fig:corner_plot}, where we see that no relevant constraints can be put on the frequency of the $(330)$ with a single event, however by combining several observations one can get interesting constraints. The results are summarized in Fig.~\ref{fig:GW150914_stacking} where we plot the 2-$\sigma$ errors for $\delta f_{220}$, $\delta f_{330}$ 
and $\delta \tau_{220}$. We find that we would need
$\sim 20$ GW150914-like events to constrain the frequency of the (220)
mode by $1\%$ at the 2-$\sigma$ level, while to constrain the damping
time of the (220) mode by $5\%$ one would need $\sim 23$ such
events. On the other hand, to constrain the frequency of the (330) by
$5\%$ we would need at least $\sim 32$ events, and we note that
this last number is highly dependent on the BBH mass ratio and inclination. From the expected rates for GW150914-like events~\cite{Abbott:2016nhf}, one can conclude that, in the best case scenario, with one year of observations at design sensitivity one could have $\mathcal{O}(10)$ GW150914-like events, therefore being able to measure $f_{330}$ with an error of the order of $10\%$ at the 2-$\sigma$ level, while $f_{220}$ and $\tau_{220}$ would be measured with 2-$\sigma$ errors of less than $2\%$ and $10\%$, respectively.

\begin{figure*}
\begin{center}
\includegraphics[width=0.48\textwidth]{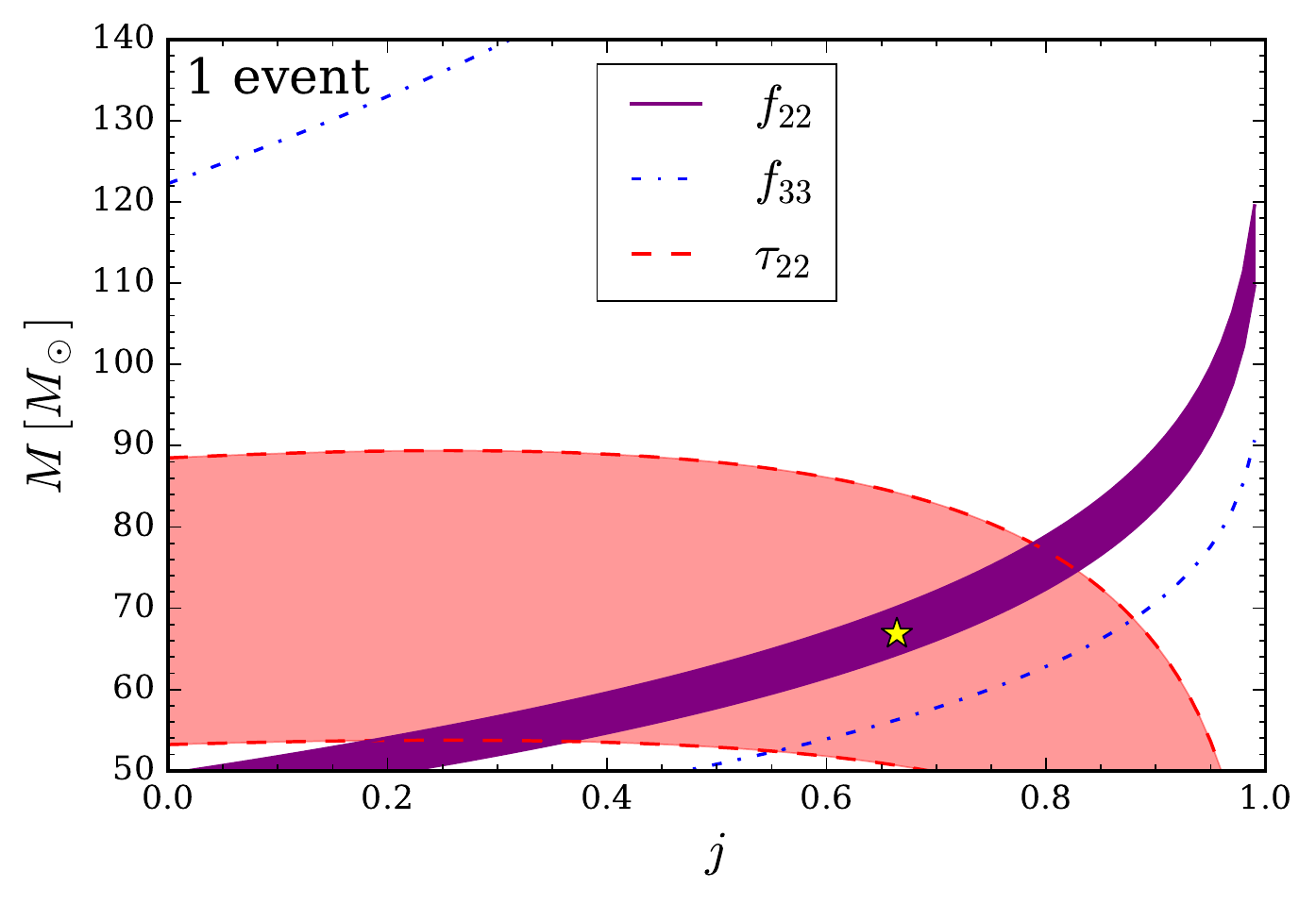}
\includegraphics[width=0.48\textwidth]{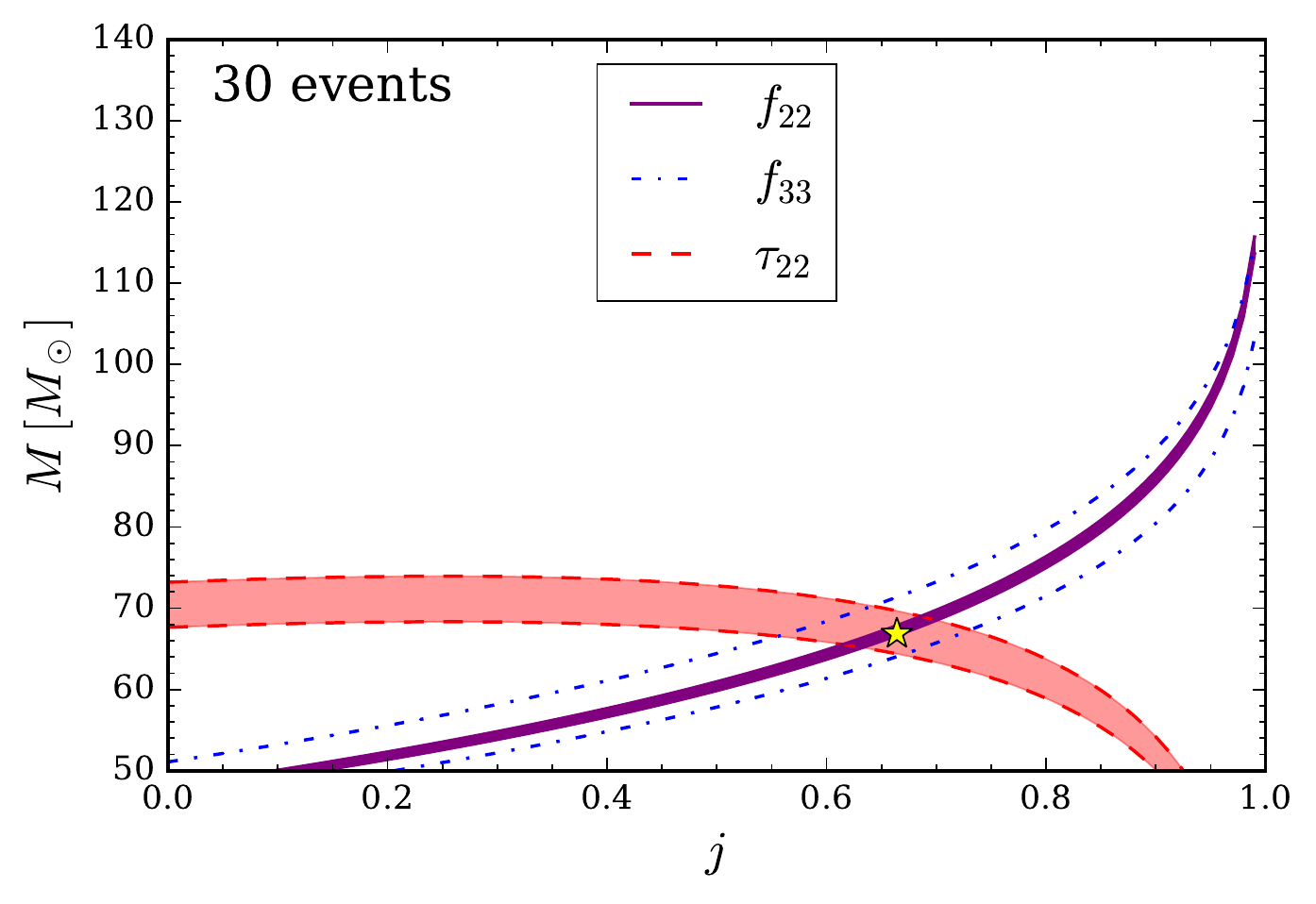}
\caption{Left: Projections of the $95\%$ confidence intervals for $(f_{22},f_{33},\tau_{22})$ in the mass ($M$) versus dimensionless spin ($j\equiv J/M^2$) of the final compact object for a GW150914-like event. The yellow star corresponds to the injected value. Right: Same, but after combining 30 GW150914-like identical events.
  \label{fig:nohairtest}}
\end{center}
\end{figure*}

A concrete way to visualize what this means in terms of testing
  the no-hair conjecture is to relate the measured QNM frequencies and
  damping times with the mass and spin of a Kerr BH in GR, as done in 
  Refs.~\cite{Kamaretsos:2011um,Gossan:2011ha}. Using the
  fits of Ref.~\cite{Berti:2005ys}, one can draw bands for each
  measured QNM parameter in the mass versus spin plane. If the bands
  do not intersect then one can invalidate GR or conclude that the
  final object is not a BH. For the GW150914-like events discussed
  above, our results are summarized in Fig.~\ref{fig:nohairtest} where
  we show the projections of the $95\%$ confidence intervals for
  $(f_{22},f_{33},\tau_{22})$ in the mass versus spin of the final
  compact object, when considering only one event and, for
  illustration, when combining $30$ identical events\footnote{We note
    that in reality all the events will have different masses and
    spins. A possible way to produce a plot similar to Fig.~\ref{fig:nohairtest} for non-identical events is to pick a reference event and use Eq.~\eqref{error} with the combined posteriors for $\delta \sigma_{lm}$ to get the BH mass and spin of the reference event. The requirement that all the bands intersect at the same point would then be equivalent to all the modes being consistent with $\delta \sigma_{lm}=0$.}. As one can see,
  the bands intersect perfectly at the injected value (yellow star in
  Fig.~\ref{fig:nohairtest}) with a significant decrease of the width
  of the bands when combining several events. This illustrates how
  combining several events can be used to severely constrain
  deviations from GR.


\section{Outlook}
\label{sec:conclusions}

We investigated the advantages of using IMR waveforms, with respect to 
damped-sinusoid models, to measure ringdown frequencies and damping times in the post-merger 
signal of a compact-object coalescence. To address this goal, we built a parameterized multipolar IMR waveform model within 
the EOB formalism (pEOBNR), and investigated its ability in measuring the QNM complex frequencies in GW150914, and 
in several synthetic GW signals injected in Gaussian noise.
 
We found the following important advantages: (i) using an IMR model, 
calibrated to NR waveforms, one does not need to define an \emph{a priori} unknown starting time at
which the signal can be described as a sum of exponentially damped
sinusoids~\cite{Gossan:2011ha,Meidam:2014jpa,Cabero:2017avf,Bhagwat:2017tkm,Baibhav:2017jhs,London:2018gaq}, therefore avoiding potential biases due to a non-optimal
choice of the ringdown starting time~\cite{Thrane:2017lqn}; (ii) the IMR model avoids technical
issues inherent to assuming a waveform with a cutoff at a particular
time, namely the need to know in advance the sky position and time at
coalescence~\cite{TheLIGOScientific:2016src,Cabero:2017avf}; (iii) the IMR model naturally includes important physics,
such as phase shifts between different modes, their relative
amplitudes and the presence of overtones~\cite{Buonanno:2006ui,Pan:2011gk}; and (iv) the IMR model
generically leads to stronger constraints on the QNM frequencies
compared to what can be achieved with a damped-sinusoid model.

The approach that we here presented should also be seen as
complementary to previous works on the subject. Besides directly
measuring the ringdown frequencies, our IMR model 
can also be used to validate the results obtained with the more agnostic
damped-sinusoid models. In particular, as we showed, the pEOBNR model already
provides very interesting constraints on the frequency and damping time of the dominant QNM of GW150914~\cite{Abbott:2016blz}.

This work can be improved in several fronts and should be seen as a
first step toward more accurate waveform models that allow to measure
deviations from GR. Although we presented results using a nonspinning BBH waveform model,
the extension to nonprecessing, spinning BBHs is straightforward
and will be done in the future, using the recently developed multipolar EOBNR 
model with spins aligned/anti-aligned with the direction perpendicular to the 
orbital plane~\cite{Cotesta:2018fcv}. Given that EOBNR models naturally encodes
time shifts between different modes and their relative amplitudes and
phases, it could in principle be used as a starting point to
perform the coherent stacking proposed in Refs.~\cite{Yang:2017zxs,DaSilvaCosta:2017njq}. A
proper implementation of the method is, however, challenging and
requires further work. The IMR model here presented could also be
extended to allow GR deviations in the inspiral phase. In addition, further work in detector noise modelling is needed 
to handle non-Gaussianities in the data. We do note that longer waveform models, such as the ones generated 
with our IMR model, are in general more robust against deviations from Gaussian noise than shorter waveform 
models, such as the damped-sinusoid models. We hope to come back to these relevant issues in the near future. Finally, we note that a MCMC parameter estimation run using the pEOBNR
model is in general much slower than using the much simpler
damped sinusoid model. However, as already done in the literature (see e.g. Ref.~\cite{Purrer:2015tud,Bohe:2016gbl}), efficient waveforms can be obtained building a reduce-order-model 
of pEOBNR waveforms, which we plan to develop in the near future.
 
\section*{Acknowledgments}

We thank Roberto Cotesta, Yuri Levin and Serguei Ossokine for useful discussions, and Michael P\"{u}rrer for a careful reading of the manuscript. We are grateful to Andrea Taracchini for help in implementing 
the parameterized multipolar effective-one-body waveform model (pEOBNR) with free (complex) QNM frequencies in the 
\textsc{LIGO Algorithm Library}. The Markov-chain Monte Carlo and Nested Sampling runs were performed on the {\sc Vulcan} cluster at the Max Planck Institute for Gravitational Physics in Potsdam.

\bibliography{Ref}

\begin{thebibliography}{78}%
\makeatletter
\providecommand \@ifxundefined [1]{%
 \@ifx{#1\undefined}
}%
\providecommand \@ifnum [1]{%
 \ifnum #1\expandafter \@firstoftwo
 \else \expandafter \@secondoftwo
 \fi
}%
\providecommand \@ifx [1]{%
 \ifx #1\expandafter \@firstoftwo
 \else \expandafter \@secondoftwo
 \fi
}%
\providecommand \natexlab [1]{#1}%
\providecommand \enquote  [1]{``#1''}%
\providecommand \bibnamefont  [1]{#1}%
\providecommand \bibfnamefont [1]{#1}%
\providecommand \citenamefont [1]{#1}%
\providecommand \href@noop [0]{\@secondoftwo}%
\providecommand \href [0]{\begingroup \@sanitize@url \@href}%
\providecommand \@href[1]{\@@startlink{#1}\@@href}%
\providecommand \@@href[1]{\endgroup#1\@@endlink}%
\providecommand \@sanitize@url [0]{\catcode `\\12\catcode `\$12\catcode
  `\&12\catcode `\#12\catcode `\^12\catcode `\_12\catcode `\%12\relax}%
\providecommand \@@startlink[1]{}%
\providecommand \@@endlink[0]{}%
\providecommand \url  [0]{\begingroup\@sanitize@url \@url }%
\providecommand \@url [1]{\endgroup\@href {#1}{\urlprefix }}%
\providecommand \urlprefix  [0]{URL }%
\providecommand \Eprint [0]{\href }%
\providecommand \doibase [0]{http://dx.doi.org/}%
\providecommand \selectlanguage [0]{\@gobble}%
\providecommand \bibinfo  [0]{\@secondoftwo}%
\providecommand \bibfield  [0]{\@secondoftwo}%
\providecommand \translation [1]{[#1]}%
\providecommand \BibitemOpen [0]{}%
\providecommand \bibitemStop [0]{}%
\providecommand \bibitemNoStop [0]{.\EOS\space}%
\providecommand \EOS [0]{\spacefactor3000\relax}%
\providecommand \BibitemShut  [1]{\csname bibitem#1\endcsname}%
\let\auto@bib@innerbib\@empty
%</preamble>
\bibitem [{\citenamefont {Abbott}\ \emph
  {et~al.}(2016{\natexlab{a}})\citenamefont {Abbott} \emph
  {et~al.}}]{Abbott:2016blz}%
  \BibitemOpen
  \bibfield  {author} {\bibinfo {author} {\bibfnamefont {B.~P.}\ \bibnamefont
  {Abbott}} \emph {et~al.} (\bibinfo {collaboration} {Virgo, LIGO
  Scientific}),\ }\href {\doibase 10.1103/PhysRevLett.116.061102} {\bibfield
  {journal} {\bibinfo  {journal} {Phys. Rev. Lett.}\ }\textbf {\bibinfo
  {volume} {116}},\ \bibinfo {pages} {061102} (\bibinfo {year}
  {2016}{\natexlab{a}})},\ \Eprint {http://arxiv.org/abs/1602.03837}
  {arXiv:1602.03837 [gr-qc]} \BibitemShut {NoStop}%
%%CITATION = ARXIV:1602.03837;%%
\bibitem [{\citenamefont {Abbott}\ \emph
  {et~al.}(2016{\natexlab{b}})\citenamefont {Abbott} \emph
  {et~al.}}]{Abbott:2016nmj}%
  \BibitemOpen
  \bibfield  {author} {\bibinfo {author} {\bibfnamefont {B.~P.}\ \bibnamefont
  {Abbott}} \emph {et~al.} (\bibinfo {collaboration} {Virgo, LIGO
  Scientific}),\ }\href {\doibase 10.1103/PhysRevLett.116.241103} {\bibfield
  {journal} {\bibinfo  {journal} {Phys. Rev. Lett.}\ }\textbf {\bibinfo
  {volume} {116}},\ \bibinfo {pages} {241103} (\bibinfo {year}
  {2016}{\natexlab{b}})},\ \Eprint {http://arxiv.org/abs/1606.04855}
  {arXiv:1606.04855 [gr-qc]} \BibitemShut {NoStop}%
%%CITATION = ARXIV:1606.04855;%%
\bibitem [{\citenamefont {Abbott}\ \emph
  {et~al.}(2017{\natexlab{a}})\citenamefont {Abbott} \emph
  {et~al.}}]{Abbott:2017vtc}%
  \BibitemOpen
  \bibfield  {author} {\bibinfo {author} {\bibfnamefont {B.~P.}\ \bibnamefont
  {Abbott}} \emph {et~al.} (\bibinfo {collaboration} {VIRGO, LIGO
  Scientific}),\ }\href {\doibase 10.1103/PhysRevLett.118.221101} {\bibfield
  {journal} {\bibinfo  {journal} {Phys. Rev. Lett.}\ }\textbf {\bibinfo
  {volume} {118}},\ \bibinfo {pages} {221101} (\bibinfo {year}
  {2017}{\natexlab{a}})},\ \Eprint {http://arxiv.org/abs/1706.01812}
  {arXiv:1706.01812 [gr-qc]} \BibitemShut {NoStop}%
%%CITATION = ARXIV:1706.01812;%%
\bibitem [{\citenamefont {Abbott}\ \emph
  {et~al.}(2017{\natexlab{b}})\citenamefont {Abbott} \emph
  {et~al.}}]{Abbott:2017gyy}%
  \BibitemOpen
  \bibfield  {author} {\bibinfo {author} {\bibfnamefont {B.~P.}\ \bibnamefont
  {Abbott}} \emph {et~al.} (\bibinfo {collaboration} {Virgo, LIGO
  Scientific}),\ }\href {\doibase 10.3847/2041-8213/aa9f0c} {\bibfield
  {journal} {\bibinfo  {journal} {Astrophys. J.}\ }\textbf {\bibinfo {volume}
  {851}},\ \bibinfo {pages} {L35} (\bibinfo {year} {2017}{\natexlab{b}})},\
  \Eprint {http://arxiv.org/abs/1711.05578} {arXiv:1711.05578 [astro-ph.HE]}
  \BibitemShut {NoStop}%
%%CITATION = ARXIV:1711.05578;%%
\bibitem [{\citenamefont {Abbott}\ \emph
  {et~al.}(2017{\natexlab{c}})\citenamefont {Abbott} \emph
  {et~al.}}]{Abbott:2017oio}%
  \BibitemOpen
  \bibfield  {author} {\bibinfo {author} {\bibfnamefont {B.~P.}\ \bibnamefont
  {Abbott}} \emph {et~al.} (\bibinfo {collaboration} {Virgo, LIGO
  Scientific}),\ }\href {\doibase 10.1103/PhysRevLett.119.141101} {\bibfield
  {journal} {\bibinfo  {journal} {Phys. Rev. Lett.}\ }\textbf {\bibinfo
  {volume} {119}},\ \bibinfo {pages} {141101} (\bibinfo {year}
  {2017}{\natexlab{c}})},\ \Eprint {http://arxiv.org/abs/1709.09660}
  {arXiv:1709.09660 [gr-qc]} \BibitemShut {NoStop}%
%%CITATION = ARXIV:1709.09660;%%
\bibitem [{\citenamefont {Abbott}\ \emph
  {et~al.}(2016{\natexlab{c}})\citenamefont {Abbott} \emph
  {et~al.}}]{TheLIGOScientific:2016src}%
  \BibitemOpen
  \bibfield  {author} {\bibinfo {author} {\bibfnamefont {B.~P.}\ \bibnamefont
  {Abbott}} \emph {et~al.} (\bibinfo {collaboration} {Virgo, LIGO
  Scientific}),\ }\href {\doibase 10.1103/PhysRevLett.116.221101} {\bibfield
  {journal} {\bibinfo  {journal} {Phys. Rev. Lett.}\ }\textbf {\bibinfo
  {volume} {116}},\ \bibinfo {pages} {221101} (\bibinfo {year}
  {2016}{\natexlab{c}})},\ \Eprint {http://arxiv.org/abs/1602.03841}
  {arXiv:1602.03841 [gr-qc]} \BibitemShut {NoStop}%
%%CITATION = ARXIV:1602.03841;%%
\bibitem [{\citenamefont {Abbott}\ \emph
  {et~al.}(2016{\natexlab{d}})\citenamefont {Abbott} \emph
  {et~al.}}]{TheLIGOScientific:2016pea}%
  \BibitemOpen
  \bibfield  {author} {\bibinfo {author} {\bibfnamefont {B.~P.}\ \bibnamefont
  {Abbott}} \emph {et~al.} (\bibinfo {collaboration} {Virgo, LIGO
  Scientific}),\ }\href {\doibase 10.1103/PhysRevX.6.041015} {\bibfield
  {journal} {\bibinfo  {journal} {Phys. Rev.}\ }\textbf {\bibinfo {volume}
  {X6}},\ \bibinfo {pages} {041015} (\bibinfo {year} {2016}{\natexlab{d}})},\
  \Eprint {http://arxiv.org/abs/1606.04856} {arXiv:1606.04856 [gr-qc]}
  \BibitemShut {NoStop}%
%%CITATION = ARXIV:1606.04856;%%
\bibitem [{\citenamefont {Abbott}\ \emph
  {et~al.}(2017{\natexlab{d}})\citenamefont {Abbott} \emph
  {et~al.}}]{TheLIGOScientific:2017qsa}%
  \BibitemOpen
  \bibfield  {author} {\bibinfo {author} {\bibfnamefont {B.~P.}\ \bibnamefont
  {Abbott}} \emph {et~al.} (\bibinfo {collaboration} {Virgo, LIGO
  Scientific}),\ }\href {\doibase 10.1103/PhysRevLett.119.161101} {\bibfield
  {journal} {\bibinfo  {journal} {Phys. Rev. Lett.}\ }\textbf {\bibinfo
  {volume} {119}},\ \bibinfo {pages} {161101} (\bibinfo {year}
  {2017}{\natexlab{d}})},\ \Eprint {http://arxiv.org/abs/1710.05832}
  {arXiv:1710.05832 [gr-qc]} \BibitemShut {NoStop}%
%%CITATION = ARXIV:1710.05832;%%
\bibitem [{\citenamefont {Abbott}\ \emph {et~al.}(2018)\citenamefont {Abbott}
  \emph {et~al.}}]{Abbott:2018wiz}%
  \BibitemOpen
  \bibfield  {author} {\bibinfo {author} {\bibfnamefont {B.~P.}\ \bibnamefont
  {Abbott}} \emph {et~al.} (\bibinfo {collaboration} {Virgo, LIGO
  Scientific}),\ }\href@noop {} {\  (\bibinfo {year} {2018})},\ \Eprint
  {http://arxiv.org/abs/1805.11579} {arXiv:1805.11579 [gr-qc]} \BibitemShut
  {NoStop}%
%%CITATION = ARXIV:1805.11579;%%
\bibitem [{\citenamefont {Kerr}(1963)}]{Kerr:1963ud}%
  \BibitemOpen
  \bibfield  {author} {\bibinfo {author} {\bibfnamefont {R.~P.}\ \bibnamefont
  {Kerr}},\ }\href {\doibase 10.1103/PhysRevLett.11.237} {\bibfield  {journal}
  {\bibinfo  {journal} {Phys. Rev. Lett.}\ }\textbf {\bibinfo {volume} {11}},\
  \bibinfo {pages} {237} (\bibinfo {year} {1963})}\BibitemShut {NoStop}%
%%CITATION = PRLTA,11,237;%%
\bibitem [{\citenamefont {Will}(2014)}]{Will:2014kxa}%
  \BibitemOpen
  \bibfield  {author} {\bibinfo {author} {\bibfnamefont {C.~M.}\ \bibnamefont
  {Will}},\ }\href {\doibase 10.12942/lrr-2014-4} {\bibfield  {journal}
  {\bibinfo  {journal} {Living Rev. Rel.}\ }\textbf {\bibinfo {volume} {17}},\
  \bibinfo {pages} {4} (\bibinfo {year} {2014})},\ \Eprint
  {http://arxiv.org/abs/1403.7377} {arXiv:1403.7377 [gr-qc]} \BibitemShut
  {NoStop}%
%%CITATION = ARXIV:1403.7377;%%
\bibitem [{\citenamefont {Berti}\ \emph {et~al.}(2015)\citenamefont {Berti}
  \emph {et~al.}}]{Berti:2015itd}%
  \BibitemOpen
  \bibfield  {author} {\bibinfo {author} {\bibfnamefont {E.}~\bibnamefont
  {Berti}} \emph {et~al.},\ }\href {\doibase 10.1088/0264-9381/32/24/243001}
  {\bibfield  {journal} {\bibinfo  {journal} {Class. Quant. Grav.}\ }\textbf
  {\bibinfo {volume} {32}},\ \bibinfo {pages} {243001} (\bibinfo {year}
  {2015})},\ \Eprint {http://arxiv.org/abs/1501.07274} {arXiv:1501.07274
  [gr-qc]} \BibitemShut {NoStop}%
%%CITATION = ARXIV:1501.07274;%%
\bibitem [{\citenamefont {Mazur}\ and\ \citenamefont
  {Mottola}(2004)}]{Mazur:2004fk}%
  \BibitemOpen
  \bibfield  {author} {\bibinfo {author} {\bibfnamefont {P.~O.}\ \bibnamefont
  {Mazur}}\ and\ \bibinfo {author} {\bibfnamefont {E.}~\bibnamefont
  {Mottola}},\ }\href {\doibase 10.1073/pnas.0402717101} {\bibfield  {journal}
  {\bibinfo  {journal} {Proc. Nat. Acad. Sci.}\ }\textbf {\bibinfo {volume}
  {101}},\ \bibinfo {pages} {9545} (\bibinfo {year} {2004})},\ \Eprint
  {http://arxiv.org/abs/gr-qc/0407075} {arXiv:gr-qc/0407075 [gr-qc]}
  \BibitemShut {NoStop}%
%%CITATION = GR-QC/0407075;%%
\bibitem [{\citenamefont {Visser}\ and\ \citenamefont
  {Wiltshire}(2004)}]{Visser:2003ge}%
  \BibitemOpen
  \bibfield  {author} {\bibinfo {author} {\bibfnamefont {M.}~\bibnamefont
  {Visser}}\ and\ \bibinfo {author} {\bibfnamefont {D.~L.}\ \bibnamefont
  {Wiltshire}},\ }\href {\doibase 10.1088/0264-9381/21/4/027} {\bibfield
  {journal} {\bibinfo  {journal} {Class. Quant. Grav.}\ }\textbf {\bibinfo
  {volume} {21}},\ \bibinfo {pages} {1135} (\bibinfo {year} {2004})},\ \Eprint
  {http://arxiv.org/abs/gr-qc/0310107} {arXiv:gr-qc/0310107 [gr-qc]}
  \BibitemShut {NoStop}%
%%CITATION = GR-QC/0310107;%%
\bibitem [{\citenamefont {Liebling}\ and\ \citenamefont
  {Palenzuela}(2012)}]{Liebling:2012fv}%
  \BibitemOpen
  \bibfield  {author} {\bibinfo {author} {\bibfnamefont {S.~L.}\ \bibnamefont
  {Liebling}}\ and\ \bibinfo {author} {\bibfnamefont {C.}~\bibnamefont
  {Palenzuela}},\ }\href {\doibase 10.12942/lrr-2012-6} {\bibfield  {journal}
  {\bibinfo  {journal} {Living Rev. Rel.}\ }\textbf {\bibinfo {volume} {15}},\
  \bibinfo {pages} {6} (\bibinfo {year} {2012})},\ \Eprint
  {http://arxiv.org/abs/1202.5809} {arXiv:1202.5809 [gr-qc]} \BibitemShut
  {NoStop}%
%%CITATION = ARXIV:1202.5809;%%
\bibitem [{\citenamefont {Vishveshwara}(1970)}]{Vishveshwara:1970zz}%
  \BibitemOpen
  \bibfield  {author} {\bibinfo {author} {\bibfnamefont {C.~V.}\ \bibnamefont
  {Vishveshwara}},\ }\href {\doibase 10.1038/227936a0} {\bibfield  {journal}
  {\bibinfo  {journal} {Nature}\ }\textbf {\bibinfo {volume} {227}},\ \bibinfo
  {pages} {936} (\bibinfo {year} {1970})}\BibitemShut {NoStop}%
%%CITATION = NATUA,227,936;%%
\bibitem [{\citenamefont {Press}(1971)}]{Press:1971wr}%
  \BibitemOpen
  \bibfield  {author} {\bibinfo {author} {\bibfnamefont {W.~H.}\ \bibnamefont
  {Press}},\ }\href {\doibase 10.1086/180849} {\bibfield  {journal} {\bibinfo
  {journal} {Astrophys. J.}\ }\textbf {\bibinfo {volume} {170}},\ \bibinfo
  {pages} {L105} (\bibinfo {year} {1971})}\BibitemShut {NoStop}%
%%CITATION = ASJOA,170,L105;%%
\bibitem [{\citenamefont {Chandrasekhar}\ and\ \citenamefont
  {Detweiler}(1975)}]{Chandrasekhar:1975zza}%
  \BibitemOpen
  \bibfield  {author} {\bibinfo {author} {\bibfnamefont {S.}~\bibnamefont
  {Chandrasekhar}}\ and\ \bibinfo {author} {\bibfnamefont {S.~L.}\ \bibnamefont
  {Detweiler}},\ }\href@noop {} {\bibfield  {journal} {\bibinfo  {journal}
  {Proc.Roy.Soc.Lond.}\ }\textbf {\bibinfo {volume} {A344}},\ \bibinfo {pages}
  {441} (\bibinfo {year} {1975})}\BibitemShut {NoStop}%
%%CITATION = PRSLA,A344,441;%%
\bibitem [{\citenamefont {Israel}(1967)}]{Israel:1967wq}%
  \BibitemOpen
  \bibfield  {author} {\bibinfo {author} {\bibfnamefont {W.}~\bibnamefont
  {Israel}},\ }\href {\doibase 10.1103/PhysRev.164.1776} {\bibfield  {journal}
  {\bibinfo  {journal} {Phys. Rev.}\ }\textbf {\bibinfo {volume} {164}},\
  \bibinfo {pages} {1776} (\bibinfo {year} {1967})}\BibitemShut {NoStop}%
%%CITATION = PHRVA,164,1776;%%
\bibitem [{\citenamefont {Carter}(1971)}]{Carter:1971zc}%
  \BibitemOpen
  \bibfield  {author} {\bibinfo {author} {\bibfnamefont {B.}~\bibnamefont
  {Carter}},\ }\href {\doibase 10.1103/PhysRevLett.26.331} {\bibfield
  {journal} {\bibinfo  {journal} {Phys. Rev. Lett.}\ }\textbf {\bibinfo
  {volume} {26}},\ \bibinfo {pages} {331} (\bibinfo {year} {1971})}\BibitemShut
  {NoStop}%
%%CITATION = PRLTA,26,331;%%
\bibitem [{\citenamefont {Hawking}(1972)}]{Hawking:1971vc}%
  \BibitemOpen
  \bibfield  {author} {\bibinfo {author} {\bibfnamefont {S.~W.}\ \bibnamefont
  {Hawking}},\ }\href {\doibase 10.1007/BF01877517} {\bibfield  {journal}
  {\bibinfo  {journal} {Commun. Math. Phys.}\ }\textbf {\bibinfo {volume}
  {25}},\ \bibinfo {pages} {152} (\bibinfo {year} {1972})}\BibitemShut
  {NoStop}%
%%CITATION = CMPHA,25,152;%%
\bibitem [{\citenamefont {Robinson}(1975)}]{Robinson:1975bv}%
  \BibitemOpen
  \bibfield  {author} {\bibinfo {author} {\bibfnamefont {D.~C.}\ \bibnamefont
  {Robinson}},\ }\href {\doibase 10.1103/PhysRevLett.34.905} {\bibfield
  {journal} {\bibinfo  {journal} {Phys. Rev. Lett.}\ }\textbf {\bibinfo
  {volume} {34}},\ \bibinfo {pages} {905} (\bibinfo {year} {1975})}\BibitemShut
  {NoStop}%
%%CITATION = PRLTA,34,905;%%
\bibitem [{\citenamefont {Dreyer}\ \emph {et~al.}(2004)\citenamefont {Dreyer},
  \citenamefont {Kelly}, \citenamefont {Krishnan}, \citenamefont {Finn},
  \citenamefont {Garrison},\ and\ \citenamefont
  {Lopez-Aleman}}]{Dreyer:2003bv}%
  \BibitemOpen
  \bibfield  {author} {\bibinfo {author} {\bibfnamefont {O.}~\bibnamefont
  {Dreyer}}, \bibinfo {author} {\bibfnamefont {B.~J.}\ \bibnamefont {Kelly}},
  \bibinfo {author} {\bibfnamefont {B.}~\bibnamefont {Krishnan}}, \bibinfo
  {author} {\bibfnamefont {L.~S.}\ \bibnamefont {Finn}}, \bibinfo {author}
  {\bibfnamefont {D.}~\bibnamefont {Garrison}}, \ and\ \bibinfo {author}
  {\bibfnamefont {R.}~\bibnamefont {Lopez-Aleman}},\ }\href {\doibase
  10.1088/0264-9381/21/4/003} {\bibfield  {journal} {\bibinfo  {journal}
  {Class. Quant. Grav.}\ }\textbf {\bibinfo {volume} {21}},\ \bibinfo {pages}
  {787} (\bibinfo {year} {2004})},\ \Eprint
  {http://arxiv.org/abs/gr-qc/0309007} {arXiv:gr-qc/0309007 [gr-qc]}
  \BibitemShut {NoStop}%
%%CITATION = GR-QC/0309007;%%
\bibitem [{\citenamefont {Berti}\ \emph {et~al.}(2006)\citenamefont {Berti},
  \citenamefont {Cardoso},\ and\ \citenamefont {Will}}]{Berti:2005ys}%
  \BibitemOpen
  \bibfield  {author} {\bibinfo {author} {\bibfnamefont {E.}~\bibnamefont
  {Berti}}, \bibinfo {author} {\bibfnamefont {V.}~\bibnamefont {Cardoso}}, \
  and\ \bibinfo {author} {\bibfnamefont {C.~M.}\ \bibnamefont {Will}},\ }\href
  {\doibase 10.1103/PhysRevD.73.064030} {\bibfield  {journal} {\bibinfo
  {journal} {Phys. Rev.}\ }\textbf {\bibinfo {volume} {D73}},\ \bibinfo {pages}
  {064030} (\bibinfo {year} {2006})},\ \Eprint
  {http://arxiv.org/abs/gr-qc/0512160} {arXiv:gr-qc/0512160 [gr-qc]}
  \BibitemShut {NoStop}%
%%CITATION = GR-QC/0512160;%%
\bibitem [{\citenamefont {Gossan}\ \emph {et~al.}(2012)\citenamefont {Gossan},
  \citenamefont {Veitch},\ and\ \citenamefont {Sathyaprakash}}]{Gossan:2011ha}%
  \BibitemOpen
  \bibfield  {author} {\bibinfo {author} {\bibfnamefont {S.}~\bibnamefont
  {Gossan}}, \bibinfo {author} {\bibfnamefont {J.}~\bibnamefont {Veitch}}, \
  and\ \bibinfo {author} {\bibfnamefont {B.~S.}\ \bibnamefont
  {Sathyaprakash}},\ }\href {\doibase 10.1103/PhysRevD.85.124056} {\bibfield
  {journal} {\bibinfo  {journal} {Phys. Rev.}\ }\textbf {\bibinfo {volume}
  {D85}},\ \bibinfo {pages} {124056} (\bibinfo {year} {2012})},\ \Eprint
  {http://arxiv.org/abs/1111.5819} {arXiv:1111.5819 [gr-qc]} \BibitemShut
  {NoStop}%
%%CITATION = ARXIV:1111.5819;%%
\bibitem [{\citenamefont {Meidam}\ \emph {et~al.}(2014)\citenamefont {Meidam},
  \citenamefont {Agathos}, \citenamefont {Van Den~Broeck}, \citenamefont
  {Veitch},\ and\ \citenamefont {Sathyaprakash}}]{Meidam:2014jpa}%
  \BibitemOpen
  \bibfield  {author} {\bibinfo {author} {\bibfnamefont {J.}~\bibnamefont
  {Meidam}}, \bibinfo {author} {\bibfnamefont {M.}~\bibnamefont {Agathos}},
  \bibinfo {author} {\bibfnamefont {C.}~\bibnamefont {Van Den~Broeck}},
  \bibinfo {author} {\bibfnamefont {J.}~\bibnamefont {Veitch}}, \ and\ \bibinfo
  {author} {\bibfnamefont {B.~S.}\ \bibnamefont {Sathyaprakash}},\ }\href
  {\doibase 10.1103/PhysRevD.90.064009} {\bibfield  {journal} {\bibinfo
  {journal} {Phys. Rev.}\ }\textbf {\bibinfo {volume} {D90}},\ \bibinfo {pages}
  {064009} (\bibinfo {year} {2014})},\ \Eprint {http://arxiv.org/abs/1406.3201}
  {arXiv:1406.3201 [gr-qc]} \BibitemShut {NoStop}%
%%CITATION = ARXIV:1406.3201;%%
\bibitem [{\citenamefont {Yang}\ \emph {et~al.}(2017)\citenamefont {Yang},
  \citenamefont {Yagi}, \citenamefont {Blackman}, \citenamefont {Lehner},
  \citenamefont {Paschalidis}, \citenamefont {Pretorius},\ and\ \citenamefont
  {Yunes}}]{Yang:2017zxs}%
  \BibitemOpen
  \bibfield  {author} {\bibinfo {author} {\bibfnamefont {H.}~\bibnamefont
  {Yang}}, \bibinfo {author} {\bibfnamefont {K.}~\bibnamefont {Yagi}}, \bibinfo
  {author} {\bibfnamefont {J.}~\bibnamefont {Blackman}}, \bibinfo {author}
  {\bibfnamefont {L.}~\bibnamefont {Lehner}}, \bibinfo {author} {\bibfnamefont
  {V.}~\bibnamefont {Paschalidis}}, \bibinfo {author} {\bibfnamefont
  {F.}~\bibnamefont {Pretorius}}, \ and\ \bibinfo {author} {\bibfnamefont
  {N.}~\bibnamefont {Yunes}},\ }\href {\doibase 10.1103/PhysRevLett.118.161101}
  {\bibfield  {journal} {\bibinfo  {journal} {Phys. Rev. Lett.}\ }\textbf
  {\bibinfo {volume} {118}},\ \bibinfo {pages} {161101} (\bibinfo {year}
  {2017})},\ \Eprint {http://arxiv.org/abs/1701.05808} {arXiv:1701.05808
  [gr-qc]} \BibitemShut {NoStop}%
%%CITATION = ARXIV:1701.05808;%%
\bibitem [{\citenamefont {Da~Silva~Costa}\ \emph {et~al.}(2017)\citenamefont
  {Da~Silva~Costa}, \citenamefont {Tiwari}, \citenamefont {Klimenko},\ and\
  \citenamefont {Salemi}}]{DaSilvaCosta:2017njq}%
  \BibitemOpen
  \bibfield  {author} {\bibinfo {author} {\bibfnamefont {C.~F.}\ \bibnamefont
  {Da~Silva~Costa}}, \bibinfo {author} {\bibfnamefont {S.}~\bibnamefont
  {Tiwari}}, \bibinfo {author} {\bibfnamefont {S.}~\bibnamefont {Klimenko}}, \
  and\ \bibinfo {author} {\bibfnamefont {F.}~\bibnamefont {Salemi}},\
  }\href@noop {} {\  (\bibinfo {year} {2017})},\ \Eprint
  {http://arxiv.org/abs/1711.00551} {arXiv:1711.00551 [gr-qc]} \BibitemShut
  {NoStop}%
%%CITATION = ARXIV:1711.00551;%%
\bibitem [{\citenamefont {Cardoso}\ and\ \citenamefont
  {Gualtieri}(2016)}]{Cardoso:2016ryw}%
  \BibitemOpen
  \bibfield  {author} {\bibinfo {author} {\bibfnamefont {V.}~\bibnamefont
  {Cardoso}}\ and\ \bibinfo {author} {\bibfnamefont {L.}~\bibnamefont
  {Gualtieri}},\ }\href {\doibase 10.1088/0264-9381/33/17/174001} {\bibfield
  {journal} {\bibinfo  {journal} {Class. Quant. Grav.}\ }\textbf {\bibinfo
  {volume} {33}},\ \bibinfo {pages} {174001} (\bibinfo {year} {2016})},\
  \Eprint {http://arxiv.org/abs/1607.03133} {arXiv:1607.03133 [gr-qc]}
  \BibitemShut {NoStop}%
%%CITATION = ARXIV:1607.03133;%%
\bibitem [{\citenamefont {Berti}\ \emph
  {et~al.}(2007{\natexlab{a}})\citenamefont {Berti}, \citenamefont {Cardoso},
  \citenamefont {Cardoso},\ and\ \citenamefont {Cavaglia}}]{Berti:2007zu}%
  \BibitemOpen
  \bibfield  {author} {\bibinfo {author} {\bibfnamefont {E.}~\bibnamefont
  {Berti}}, \bibinfo {author} {\bibfnamefont {J.}~\bibnamefont {Cardoso}},
  \bibinfo {author} {\bibfnamefont {V.}~\bibnamefont {Cardoso}}, \ and\
  \bibinfo {author} {\bibfnamefont {M.}~\bibnamefont {Cavaglia}},\ }\href
  {\doibase 10.1103/PhysRevD.76.104044} {\bibfield  {journal} {\bibinfo
  {journal} {Phys. Rev.}\ }\textbf {\bibinfo {volume} {D76}},\ \bibinfo {pages}
  {104044} (\bibinfo {year} {2007}{\natexlab{a}})},\ \Eprint
  {http://arxiv.org/abs/0707.1202} {arXiv:0707.1202 [gr-qc]} \BibitemShut
  {NoStop}%
%%CITATION = ARXIV:0707.1202;%%
\bibitem [{\citenamefont {Bhagwat}\ \emph {et~al.}(2016)\citenamefont
  {Bhagwat}, \citenamefont {Brown},\ and\ \citenamefont
  {Ballmer}}]{Bhagwat:2016ntk}%
  \BibitemOpen
  \bibfield  {author} {\bibinfo {author} {\bibfnamefont {S.}~\bibnamefont
  {Bhagwat}}, \bibinfo {author} {\bibfnamefont {D.~A.}\ \bibnamefont {Brown}},
  \ and\ \bibinfo {author} {\bibfnamefont {S.~W.}\ \bibnamefont {Ballmer}},\
  }\href {\doibase 10.1103/PhysRevD.94.084024, 10.1103/PhysRevD.95.069906}
  {\bibfield  {journal} {\bibinfo  {journal} {Phys. Rev.}\ }\textbf {\bibinfo
  {volume} {D94}},\ \bibinfo {pages} {084024} (\bibinfo {year} {2016})},\
  \bibinfo {note} {[Erratum: Phys. Rev.D95,no.6,069906(2017)]},\ \Eprint
  {http://arxiv.org/abs/1607.07845} {arXiv:1607.07845 [gr-qc]} \BibitemShut
  {NoStop}%
%%CITATION = ARXIV:1607.07845;%%
\bibitem [{\citenamefont {Berti}\ \emph {et~al.}(2016)\citenamefont {Berti},
  \citenamefont {Sesana}, \citenamefont {Barausse}, \citenamefont {Cardoso},\
  and\ \citenamefont {Belczynski}}]{Berti:2016lat}%
  \BibitemOpen
  \bibfield  {author} {\bibinfo {author} {\bibfnamefont {E.}~\bibnamefont
  {Berti}}, \bibinfo {author} {\bibfnamefont {A.}~\bibnamefont {Sesana}},
  \bibinfo {author} {\bibfnamefont {E.}~\bibnamefont {Barausse}}, \bibinfo
  {author} {\bibfnamefont {V.}~\bibnamefont {Cardoso}}, \ and\ \bibinfo
  {author} {\bibfnamefont {K.}~\bibnamefont {Belczynski}},\ }\href {\doibase
  10.1103/PhysRevLett.117.101102} {\bibfield  {journal} {\bibinfo  {journal}
  {Phys. Rev. Lett.}\ }\textbf {\bibinfo {volume} {117}},\ \bibinfo {pages}
  {101102} (\bibinfo {year} {2016})},\ \Eprint
  {http://arxiv.org/abs/1605.09286} {arXiv:1605.09286 [gr-qc]} \BibitemShut
  {NoStop}%
%%CITATION = ARXIV:1605.09286;%%
\bibitem [{\citenamefont {Maselli}\ \emph {et~al.}(2017)\citenamefont
  {Maselli}, \citenamefont {Kokkotas},\ and\ \citenamefont
  {Laguna}}]{Maselli:2017kvl}%
  \BibitemOpen
  \bibfield  {author} {\bibinfo {author} {\bibfnamefont {A.}~\bibnamefont
  {Maselli}}, \bibinfo {author} {\bibfnamefont {K.}~\bibnamefont {Kokkotas}}, \
  and\ \bibinfo {author} {\bibfnamefont {P.}~\bibnamefont {Laguna}},\ }\href
  {\doibase 10.1103/PhysRevD.95.104026} {\bibfield  {journal} {\bibinfo
  {journal} {Phys. Rev.}\ }\textbf {\bibinfo {volume} {D95}},\ \bibinfo {pages}
  {104026} (\bibinfo {year} {2017})},\ \Eprint
  {http://arxiv.org/abs/1702.01110} {arXiv:1702.01110 [gr-qc]} \BibitemShut
  {NoStop}%
%%CITATION = ARXIV:1702.01110;%%
\bibitem [{\citenamefont {Cabero}\ \emph {et~al.}(2017)\citenamefont {Cabero},
  \citenamefont {Capano}, \citenamefont {Fischer-Birnholtz}, \citenamefont
  {Krishnan}, \citenamefont {Nielsen},\ and\ \citenamefont
  {Nitz}}]{Cabero:2017avf}%
  \BibitemOpen
  \bibfield  {author} {\bibinfo {author} {\bibfnamefont {M.}~\bibnamefont
  {Cabero}}, \bibinfo {author} {\bibfnamefont {C.~D.}\ \bibnamefont {Capano}},
  \bibinfo {author} {\bibfnamefont {O.}~\bibnamefont {Fischer-Birnholtz}},
  \bibinfo {author} {\bibfnamefont {B.}~\bibnamefont {Krishnan}}, \bibinfo
  {author} {\bibfnamefont {A.~B.}\ \bibnamefont {Nielsen}}, \ and\ \bibinfo
  {author} {\bibfnamefont {A.~H.}\ \bibnamefont {Nitz}},\ }\href@noop {} {\
  (\bibinfo {year} {2017})},\ \Eprint {http://arxiv.org/abs/1711.09073}
  {arXiv:1711.09073 [gr-qc]} \BibitemShut {NoStop}%
%%CITATION = ARXIV:1711.09073;%%
\bibitem [{\citenamefont {Buonanno}\ and\ \citenamefont
  {Damour}(1999)}]{Buonanno:1998gg}%
  \BibitemOpen
  \bibfield  {author} {\bibinfo {author} {\bibfnamefont {A.}~\bibnamefont
  {Buonanno}}\ and\ \bibinfo {author} {\bibfnamefont {T.}~\bibnamefont
  {Damour}},\ }\href {\doibase 10.1103/PhysRevD.59.084006} {\bibfield
  {journal} {\bibinfo  {journal} {Phys. Rev.}\ }\textbf {\bibinfo {volume}
  {D59}},\ \bibinfo {pages} {084006} (\bibinfo {year} {1999})},\ \Eprint
  {http://arxiv.org/abs/gr-qc/9811091} {arXiv:gr-qc/9811091 [gr-qc]}
  \BibitemShut {NoStop}%
%%CITATION = GR-QC/9811091;%%
\bibitem [{\citenamefont {Buonanno}\ and\ \citenamefont
  {Damour}(2000)}]{Buonanno:2000ef}%
  \BibitemOpen
  \bibfield  {author} {\bibinfo {author} {\bibfnamefont {A.}~\bibnamefont
  {Buonanno}}\ and\ \bibinfo {author} {\bibfnamefont {T.}~\bibnamefont
  {Damour}},\ }\href {\doibase 10.1103/PhysRevD.62.064015} {\bibfield
  {journal} {\bibinfo  {journal} {Phys. Rev.}\ }\textbf {\bibinfo {volume}
  {D62}},\ \bibinfo {pages} {064015} (\bibinfo {year} {2000})},\ \Eprint
  {http://arxiv.org/abs/gr-qc/0001013} {arXiv:gr-qc/0001013 [gr-qc]}
  \BibitemShut {NoStop}%
%%CITATION = GR-QC/0001013;%%
\bibitem [{\citenamefont {Pan}\ \emph {et~al.}(2011)\citenamefont {Pan},
  \citenamefont {Buonanno}, \citenamefont {Boyle}, \citenamefont {Buchman},
  \citenamefont {Kidder}, \citenamefont {Pfeiffer},\ and\ \citenamefont
  {Scheel}}]{Pan:2011gk}%
  \BibitemOpen
  \bibfield  {author} {\bibinfo {author} {\bibfnamefont {Y.}~\bibnamefont
  {Pan}}, \bibinfo {author} {\bibfnamefont {A.}~\bibnamefont {Buonanno}},
  \bibinfo {author} {\bibfnamefont {M.}~\bibnamefont {Boyle}}, \bibinfo
  {author} {\bibfnamefont {L.~T.}\ \bibnamefont {Buchman}}, \bibinfo {author}
  {\bibfnamefont {L.~E.}\ \bibnamefont {Kidder}}, \bibinfo {author}
  {\bibfnamefont {H.~P.}\ \bibnamefont {Pfeiffer}}, \ and\ \bibinfo {author}
  {\bibfnamefont {M.~A.}\ \bibnamefont {Scheel}},\ }\href {\doibase
  10.1103/PhysRevD.84.124052} {\bibfield  {journal} {\bibinfo  {journal} {Phys.
  Rev.}\ }\textbf {\bibinfo {volume} {D84}},\ \bibinfo {pages} {124052}
  (\bibinfo {year} {2011})},\ \Eprint {http://arxiv.org/abs/1106.1021}
  {arXiv:1106.1021 [gr-qc]} \BibitemShut {NoStop}%
%%CITATION = ARXIV:1106.1021;%%
\bibitem [{\citenamefont {Thrane}\ \emph {et~al.}(2017)\citenamefont {Thrane},
  \citenamefont {Lasky},\ and\ \citenamefont {Levin}}]{Thrane:2017lqn}%
  \BibitemOpen
  \bibfield  {author} {\bibinfo {author} {\bibfnamefont {E.}~\bibnamefont
  {Thrane}}, \bibinfo {author} {\bibfnamefont {P.~D.}\ \bibnamefont {Lasky}}, \
  and\ \bibinfo {author} {\bibfnamefont {Y.}~\bibnamefont {Levin}},\ }\href
  {\doibase 10.1103/PhysRevD.96.102004} {\bibfield  {journal} {\bibinfo
  {journal} {Phys. Rev.}\ }\textbf {\bibinfo {volume} {D96}},\ \bibinfo {pages}
  {102004} (\bibinfo {year} {2017})},\ \Eprint
  {http://arxiv.org/abs/1706.05152} {arXiv:1706.05152 [gr-qc]} \BibitemShut
  {NoStop}%
%%CITATION = ARXIV:1706.05152;%%
\bibitem [{\citenamefont {Barausse}\ \emph {et~al.}(2012)\citenamefont
  {Barausse}, \citenamefont {Buonanno}, \citenamefont {Hughes}, \citenamefont
  {Khanna}, \citenamefont {O'Sullivan},\ and\ \citenamefont
  {Pan}}]{Barausse:2011kb}%
  \BibitemOpen
  \bibfield  {author} {\bibinfo {author} {\bibfnamefont {E.}~\bibnamefont
  {Barausse}}, \bibinfo {author} {\bibfnamefont {A.}~\bibnamefont {Buonanno}},
  \bibinfo {author} {\bibfnamefont {S.~A.}\ \bibnamefont {Hughes}}, \bibinfo
  {author} {\bibfnamefont {G.}~\bibnamefont {Khanna}}, \bibinfo {author}
  {\bibfnamefont {S.}~\bibnamefont {O'Sullivan}}, \ and\ \bibinfo {author}
  {\bibfnamefont {Y.}~\bibnamefont {Pan}},\ }\href {\doibase
  10.1103/PhysRevD.85.024046} {\bibfield  {journal} {\bibinfo  {journal} {Phys.
  Rev.}\ }\textbf {\bibinfo {volume} {D85}},\ \bibinfo {pages} {024046}
  (\bibinfo {year} {2012})},\ \Eprint {http://arxiv.org/abs/1110.3081}
  {arXiv:1110.3081 [gr-qc]} \BibitemShut {NoStop}%
%%CITATION = ARXIV:1110.3081;%%
\bibitem [{\citenamefont {Mroue}\ \emph {et~al.}(2013)\citenamefont {Mroue}
  \emph {et~al.}}]{Mroue:2013xna}%
  \BibitemOpen
  \bibfield  {author} {\bibinfo {author} {\bibfnamefont {A.~H.}\ \bibnamefont
  {Mroue}} \emph {et~al.},\ }\href {\doibase 10.1103/PhysRevLett.111.241104}
  {\bibfield  {journal} {\bibinfo  {journal} {Phys. Rev. Lett.}\ }\textbf
  {\bibinfo {volume} {111}},\ \bibinfo {pages} {241104} (\bibinfo {year}
  {2013})},\ \Eprint {http://arxiv.org/abs/1304.6077} {arXiv:1304.6077 [gr-qc]}
  \BibitemShut {NoStop}%
%%CITATION = ARXIV:1304.6077;%%
\bibitem [{\citenamefont {Juli\'{e}}\ and\ \citenamefont
  {Deruelle}(2017)}]{Julie:2017pkb}%
  \BibitemOpen
  \bibfield  {author} {\bibinfo {author} {\bibfnamefont {F.-L.}\ \bibnamefont
  {Juli\'{e}}}\ and\ \bibinfo {author} {\bibfnamefont {N.}~\bibnamefont
  {Deruelle}},\ }\href {\doibase 10.1103/PhysRevD.95.124054} {\bibfield
  {journal} {\bibinfo  {journal} {Phys. Rev.}\ }\textbf {\bibinfo {volume}
  {D95}},\ \bibinfo {pages} {124054} (\bibinfo {year} {2017})},\ \Eprint
  {http://arxiv.org/abs/1703.05360} {arXiv:1703.05360 [gr-qc]} \BibitemShut
  {NoStop}%
%%CITATION = ARXIV:1703.05360;%%
\bibitem [{\citenamefont {Juli\'{e}}(2018)}]{Julie:2017ucp}%
  \BibitemOpen
  \bibfield  {author} {\bibinfo {author} {\bibfnamefont {F.-L.}\ \bibnamefont
  {Juli\'{e}}},\ }\href {\doibase 10.1103/PhysRevD.97.024047} {\bibfield
  {journal} {\bibinfo  {journal} {Phys. Rev.}\ }\textbf {\bibinfo {volume}
  {D97}},\ \bibinfo {pages} {024047} (\bibinfo {year} {2018})},\ \Eprint
  {http://arxiv.org/abs/1709.09742} {arXiv:1709.09742 [gr-qc]} \BibitemShut
  {NoStop}%
%%CITATION = ARXIV:1709.09742;%%
\bibitem [{\citenamefont {Kamaretsos}\ \emph {et~al.}(2012)\citenamefont
  {Kamaretsos}, \citenamefont {Hannam}, \citenamefont {Husa},\ and\
  \citenamefont {Sathyaprakash}}]{Kamaretsos:2011um}%
  \BibitemOpen
  \bibfield  {author} {\bibinfo {author} {\bibfnamefont {I.}~\bibnamefont
  {Kamaretsos}}, \bibinfo {author} {\bibfnamefont {M.}~\bibnamefont {Hannam}},
  \bibinfo {author} {\bibfnamefont {S.}~\bibnamefont {Husa}}, \ and\ \bibinfo
  {author} {\bibfnamefont {B.~S.}\ \bibnamefont {Sathyaprakash}},\ }\href
  {\doibase 10.1103/PhysRevD.85.024018} {\bibfield  {journal} {\bibinfo
  {journal} {Phys. Rev.}\ }\textbf {\bibinfo {volume} {D85}},\ \bibinfo {pages}
  {024018} (\bibinfo {year} {2012})},\ \Eprint {http://arxiv.org/abs/1107.0854}
  {arXiv:1107.0854 [gr-qc]} \BibitemShut {NoStop}%
%%CITATION = ARXIV:1107.0854;%%
\bibitem [{\citenamefont {London}\ \emph {et~al.}(2014)\citenamefont {London},
  \citenamefont {Shoemaker},\ and\ \citenamefont {Healy}}]{London:2014cma}%
  \BibitemOpen
  \bibfield  {author} {\bibinfo {author} {\bibfnamefont {L.}~\bibnamefont
  {London}}, \bibinfo {author} {\bibfnamefont {D.}~\bibnamefont {Shoemaker}}, \
  and\ \bibinfo {author} {\bibfnamefont {J.}~\bibnamefont {Healy}},\ }\href
  {\doibase 10.1103/PhysRevD.90.124032, 10.1103/PhysRevD.94.069902} {\bibfield
  {journal} {\bibinfo  {journal} {Phys. Rev.}\ }\textbf {\bibinfo {volume}
  {D90}},\ \bibinfo {pages} {124032} (\bibinfo {year} {2014})},\ \bibinfo
  {note} {[Erratum: Phys. Rev.D94,no.6,069902(2016)]},\ \Eprint
  {http://arxiv.org/abs/1404.3197} {arXiv:1404.3197 [gr-qc]} \BibitemShut
  {NoStop}%
%%CITATION = ARXIV:1404.3197;%%
\bibitem [{\citenamefont {London}(2018)}]{London:2018gaq}%
  \BibitemOpen
  \bibfield  {author} {\bibinfo {author} {\bibfnamefont {L.~T.}\ \bibnamefont
  {London}},\ }\href@noop {} {\  (\bibinfo {year} {2018})},\ \Eprint
  {http://arxiv.org/abs/1801.08208} {arXiv:1801.08208 [gr-qc]} \BibitemShut
  {NoStop}%
%%CITATION = ARXIV:1801.08208;%%
\bibitem [{\citenamefont {Buonanno}\ \emph {et~al.}(2007)\citenamefont
  {Buonanno}, \citenamefont {Cook},\ and\ \citenamefont
  {Pretorius}}]{Buonanno:2006ui}%
  \BibitemOpen
  \bibfield  {author} {\bibinfo {author} {\bibfnamefont {A.}~\bibnamefont
  {Buonanno}}, \bibinfo {author} {\bibfnamefont {G.~B.}\ \bibnamefont {Cook}},
  \ and\ \bibinfo {author} {\bibfnamefont {F.}~\bibnamefont {Pretorius}},\
  }\href {\doibase 10.1103/PhysRevD.75.124018} {\bibfield  {journal} {\bibinfo
  {journal} {Phys. Rev.}\ }\textbf {\bibinfo {volume} {D75}},\ \bibinfo {pages}
  {124018} (\bibinfo {year} {2007})},\ \Eprint
  {http://arxiv.org/abs/gr-qc/0610122} {arXiv:gr-qc/0610122 [gr-qc]}
  \BibitemShut {NoStop}%
%%CITATION = GR-QC/0610122;%%
\bibitem [{\citenamefont {Berti}\ \emph
  {et~al.}(2007{\natexlab{b}})\citenamefont {Berti}, \citenamefont {Cardoso},
  \citenamefont {Gonzalez}, \citenamefont {Sperhake}, \citenamefont {Hannam},
  \citenamefont {Husa},\ and\ \citenamefont {Bruegmann}}]{Berti:2007fi}%
  \BibitemOpen
  \bibfield  {author} {\bibinfo {author} {\bibfnamefont {E.}~\bibnamefont
  {Berti}}, \bibinfo {author} {\bibfnamefont {V.}~\bibnamefont {Cardoso}},
  \bibinfo {author} {\bibfnamefont {J.~A.}\ \bibnamefont {Gonzalez}}, \bibinfo
  {author} {\bibfnamefont {U.}~\bibnamefont {Sperhake}}, \bibinfo {author}
  {\bibfnamefont {M.}~\bibnamefont {Hannam}}, \bibinfo {author} {\bibfnamefont
  {S.}~\bibnamefont {Husa}}, \ and\ \bibinfo {author} {\bibfnamefont
  {B.}~\bibnamefont {Bruegmann}},\ }\href {\doibase 10.1103/PhysRevD.76.064034}
  {\bibfield  {journal} {\bibinfo  {journal} {Phys. Rev.}\ }\textbf {\bibinfo
  {volume} {D76}},\ \bibinfo {pages} {064034} (\bibinfo {year}
  {2007}{\natexlab{b}})},\ \Eprint {http://arxiv.org/abs/gr-qc/0703053}
  {arXiv:gr-qc/0703053 [GR-QC]} \BibitemShut {NoStop}%
%%CITATION = GR-QC/0703053;%%
\bibitem [{\citenamefont {Bayes}\ and\ \citenamefont
  {Price}(1763)}]{Bayes:1793}%
  \BibitemOpen
  \bibfield  {author} {\bibinfo {author} {\bibfnamefont {T.}~\bibnamefont
  {Bayes}}\ and\ \bibinfo {author} {\bibfnamefont {R.}~\bibnamefont {Price}},\
  }\href {\doibase 10.1098/rstl.1763.0053} {\bibfield  {journal} {\bibinfo
  {journal} {Phil. Trans. Roy. Soc. Lond.}\ }\textbf {\bibinfo {volume} {53}},\
  \bibinfo {pages} {370} (\bibinfo {year} {1763})}\BibitemShut {NoStop}%
\bibitem [{\citenamefont {Jaynes}(2003)}]{Jaynes:2003}%
  \BibitemOpen
  \bibfield  {author} {\bibinfo {author} {\bibfnamefont {E.~T.}\ \bibnamefont
  {Jaynes}},\ }\href@noop {} {\emph {\bibinfo {title} {{Probability Theory: The
  Logic of Science}}}},\ edited by\ \bibinfo {editor} {\bibfnamefont {G.~L.}\
  \bibnamefont {Bretthorst}}\ (\bibinfo  {publisher} {Cambridge University
  Press},\ \bibinfo {address} {Cambridge},\ \bibinfo {year} {2003})\BibitemShut
  {NoStop}%
\bibitem [{\citenamefont {Veitch}\ \emph {et~al.}(2015)\citenamefont {Veitch}
  \emph {et~al.}}]{veitch:2014wba}%
  \BibitemOpen
  \bibfield  {author} {\bibinfo {author} {\bibfnamefont {J.}~\bibnamefont
  {Veitch}} \emph {et~al.},\ }\href {\doibase 10.1103/PhysRevD.91.042003}
  {\bibfield  {journal} {\bibinfo  {journal} {Phys.Rev.}\ }\textbf {\bibinfo
  {volume} {D91}},\ \bibinfo {pages} {042003} (\bibinfo {year} {2015})},\
  \Eprint {http://arxiv.org/abs/1409.7215} {arXiv:1409.7215 [gr-qc]}
  \BibitemShut {NoStop}%
%%CITATION = ARXIV:1409.7215;%%
\bibitem [{\citenamefont {Finn}(1992)}]{Finn:1992wt}%
  \BibitemOpen
  \bibfield  {author} {\bibinfo {author} {\bibfnamefont {L.~S.}\ \bibnamefont
  {Finn}},\ }\href {\doibase 10.1103/PhysRevD.46.5236} {\bibfield  {journal}
  {\bibinfo  {journal} {Phys. Rev.}\ }\textbf {\bibinfo {volume} {D46}},\
  \bibinfo {pages} {5236} (\bibinfo {year} {1992})},\ \Eprint
  {http://arxiv.org/abs/gr-qc/9209010} {arXiv:gr-qc/9209010 [gr-qc]}
  \BibitemShut {NoStop}%
%%CITATION = GR-QC/9209010;%%
\bibitem [{\citenamefont {Shoemaker}(2010)}]{Shoemaker:2010}%
  \BibitemOpen
  \bibfield  {author} {\bibinfo {author} {\bibfnamefont {D.}~\bibnamefont
  {Shoemaker}} (\bibinfo {collaboration} {{LIGO} Collaboration}),\ }\href
  {https://dcc.ligo.org/cgi-bin/DocDB/ShowDocument?docid=2974} {\enquote
  {\bibinfo {title} {Advanced {LIGO} anticipated sensitivity curves},}\ }
  (\bibinfo {year} {2010}),\ \bibinfo {note} {{LIGO} Document
  T0900288-v3}\BibitemShut {NoStop}%
\bibitem [{\citenamefont {{Manzotti}}\ and\ \citenamefont
  {{Dietz}}(2012)}]{2012arXiv1202.4031M}%
  \BibitemOpen
  \bibfield  {author} {\bibinfo {author} {\bibfnamefont {A.}~\bibnamefont
  {{Manzotti}}}\ and\ \bibinfo {author} {\bibfnamefont {A.}~\bibnamefont
  {{Dietz}}},\ }\href@noop {} {\bibfield  {journal} {\bibinfo  {journal} {ArXiv
  e-prints}\ } (\bibinfo {year} {2012})},\ \Eprint
  {http://arxiv.org/abs/1202.4031} {arXiv:1202.4031 [gr-qc]} \BibitemShut
  {NoStop}%
\bibitem [{\citenamefont {Abbott}\ \emph
  {et~al.}(2016{\natexlab{e}})\citenamefont {Abbott} \emph
  {et~al.}}]{TheLIGOScientific:2016wfe}%
  \BibitemOpen
  \bibfield  {author} {\bibinfo {author} {\bibfnamefont {B.~P.}\ \bibnamefont
  {Abbott}} \emph {et~al.} (\bibinfo {collaboration} {Virgo, LIGO
  Scientific}),\ }\href {\doibase 10.1103/PhysRevLett.116.241102} {\bibfield
  {journal} {\bibinfo  {journal} {Phys. Rev. Lett.}\ }\textbf {\bibinfo
  {volume} {116}},\ \bibinfo {pages} {241102} (\bibinfo {year}
  {2016}{\natexlab{e}})},\ \Eprint {http://arxiv.org/abs/1602.03840}
  {arXiv:1602.03840 [gr-qc]} \BibitemShut {NoStop}%
%%CITATION = ARXIV:1602.03840;%%
\bibitem [{\citenamefont {Baibhav}\ \emph {et~al.}(2018)\citenamefont
  {Baibhav}, \citenamefont {Berti}, \citenamefont {Cardoso},\ and\
  \citenamefont {Khanna}}]{Baibhav:2017jhs}%
  \BibitemOpen
  \bibfield  {author} {\bibinfo {author} {\bibfnamefont {V.}~\bibnamefont
  {Baibhav}}, \bibinfo {author} {\bibfnamefont {E.}~\bibnamefont {Berti}},
  \bibinfo {author} {\bibfnamefont {V.}~\bibnamefont {Cardoso}}, \ and\
  \bibinfo {author} {\bibfnamefont {G.}~\bibnamefont {Khanna}},\ }\href
  {\doibase 10.1103/PhysRevD.97.044048} {\bibfield  {journal} {\bibinfo
  {journal} {Phys. Rev.}\ }\textbf {\bibinfo {volume} {D97}},\ \bibinfo {pages}
  {044048} (\bibinfo {year} {2018})},\ \Eprint
  {http://arxiv.org/abs/1710.02156} {arXiv:1710.02156 [gr-qc]} \BibitemShut
  {NoStop}%
%%CITATION = ARXIV:1710.02156;%%
\bibitem [{\citenamefont {Bhagwat}\ \emph {et~al.}(2017)\citenamefont
  {Bhagwat}, \citenamefont {Okounkova}, \citenamefont {Ballmer}, \citenamefont
  {Brown}, \citenamefont {Giesler}, \citenamefont {Scheel},\ and\ \citenamefont
  {Teukolsky}}]{Bhagwat:2017tkm}%
  \BibitemOpen
  \bibfield  {author} {\bibinfo {author} {\bibfnamefont {S.}~\bibnamefont
  {Bhagwat}}, \bibinfo {author} {\bibfnamefont {M.}~\bibnamefont {Okounkova}},
  \bibinfo {author} {\bibfnamefont {S.~W.}\ \bibnamefont {Ballmer}}, \bibinfo
  {author} {\bibfnamefont {D.~A.}\ \bibnamefont {Brown}}, \bibinfo {author}
  {\bibfnamefont {M.}~\bibnamefont {Giesler}}, \bibinfo {author} {\bibfnamefont
  {M.~A.}\ \bibnamefont {Scheel}}, \ and\ \bibinfo {author} {\bibfnamefont
  {S.~A.}\ \bibnamefont {Teukolsky}},\ }\href@noop {} {\  (\bibinfo {year}
  {2017})},\ \Eprint {http://arxiv.org/abs/1711.00926} {arXiv:1711.00926
  [gr-qc]} \BibitemShut {NoStop}%
%%CITATION = ARXIV:1711.00926;%%
\bibitem [{\citenamefont {Littenberg}\ \emph {et~al.}(2013)\citenamefont
  {Littenberg}, \citenamefont {Baker}, \citenamefont {Buonanno},\ and\
  \citenamefont {Kelly}}]{Littenberg:2012uj}%
  \BibitemOpen
  \bibfield  {author} {\bibinfo {author} {\bibfnamefont {T.~B.}\ \bibnamefont
  {Littenberg}}, \bibinfo {author} {\bibfnamefont {J.~G.}\ \bibnamefont
  {Baker}}, \bibinfo {author} {\bibfnamefont {A.}~\bibnamefont {Buonanno}}, \
  and\ \bibinfo {author} {\bibfnamefont {B.~J.}\ \bibnamefont {Kelly}},\ }\href
  {\doibase 10.1103/PhysRevD.87.104003} {\bibfield  {journal} {\bibinfo
  {journal} {Phys. Rev.}\ }\textbf {\bibinfo {volume} {D87}},\ \bibinfo {pages}
  {104003} (\bibinfo {year} {2013})},\ \Eprint {http://arxiv.org/abs/1210.0893}
  {arXiv:1210.0893 [gr-qc]} \BibitemShut {NoStop}%
%%CITATION = ARXIV:1210.0893;%%
\bibitem [{\citenamefont {Cotesta}\ \emph {et~al.}(2018)\citenamefont
  {Cotesta}, \citenamefont {Buonanno}, \citenamefont {Bohe}, \citenamefont
  {Taracchini}, \citenamefont {Hinder},\ and\ \citenamefont
  {Ossokine}}]{Cotesta:2018fcv}%
  \BibitemOpen
  \bibfield  {author} {\bibinfo {author} {\bibfnamefont {R.}~\bibnamefont
  {Cotesta}}, \bibinfo {author} {\bibfnamefont {A.}~\bibnamefont {Buonanno}},
  \bibinfo {author} {\bibfnamefont {A.}~\bibnamefont {Bohe}}, \bibinfo {author}
  {\bibfnamefont {A.}~\bibnamefont {Taracchini}}, \bibinfo {author}
  {\bibfnamefont {I.}~\bibnamefont {Hinder}}, \ and\ \bibinfo {author}
  {\bibfnamefont {S.}~\bibnamefont {Ossokine}},\ }\href@noop {} {\  (\bibinfo
  {year} {2018})},\ \Eprint {http://arxiv.org/abs/1803.10701} {arXiv:1803.10701
  [gr-qc]} \BibitemShut {NoStop}%
%%CITATION = ARXIV:1803.10701;%%
\bibitem [{\citenamefont {Del~Pozzo}\ \emph {et~al.}(2011)\citenamefont
  {Del~Pozzo}, \citenamefont {Veitch},\ and\ \citenamefont
  {Vecchio}}]{DelPozzo:2011pg}%
  \BibitemOpen
  \bibfield  {author} {\bibinfo {author} {\bibfnamefont {W.}~\bibnamefont
  {Del~Pozzo}}, \bibinfo {author} {\bibfnamefont {J.}~\bibnamefont {Veitch}}, \
  and\ \bibinfo {author} {\bibfnamefont {A.}~\bibnamefont {Vecchio}},\ }\href
  {\doibase 10.1103/PhysRevD.83.082002} {\bibfield  {journal} {\bibinfo
  {journal} {Phys. Rev.}\ }\textbf {\bibinfo {volume} {D83}},\ \bibinfo {pages}
  {082002} (\bibinfo {year} {2011})},\ \Eprint {http://arxiv.org/abs/1101.1391}
  {arXiv:1101.1391 [gr-qc]} \BibitemShut {NoStop}%
%%CITATION = ARXIV:1101.1391;%%
\bibitem [{\citenamefont {Li}\ \emph {et~al.}(2012)\citenamefont {Li},
  \citenamefont {Del~Pozzo}, \citenamefont {Vitale}, \citenamefont {Van
  Den~Broeck}, \citenamefont {Agathos}, \citenamefont {Veitch}, \citenamefont
  {Grover}, \citenamefont {Sidery}, \citenamefont {Sturani},\ and\
  \citenamefont {Vecchio}}]{Li:2011cg}%
  \BibitemOpen
  \bibfield  {author} {\bibinfo {author} {\bibfnamefont {T.~G.~F.}\
  \bibnamefont {Li}}, \bibinfo {author} {\bibfnamefont {W.}~\bibnamefont
  {Del~Pozzo}}, \bibinfo {author} {\bibfnamefont {S.}~\bibnamefont {Vitale}},
  \bibinfo {author} {\bibfnamefont {C.}~\bibnamefont {Van Den~Broeck}},
  \bibinfo {author} {\bibfnamefont {M.}~\bibnamefont {Agathos}}, \bibinfo
  {author} {\bibfnamefont {J.}~\bibnamefont {Veitch}}, \bibinfo {author}
  {\bibfnamefont {K.}~\bibnamefont {Grover}}, \bibinfo {author} {\bibfnamefont
  {T.}~\bibnamefont {Sidery}}, \bibinfo {author} {\bibfnamefont
  {R.}~\bibnamefont {Sturani}}, \ and\ \bibinfo {author} {\bibfnamefont
  {A.}~\bibnamefont {Vecchio}},\ }\href {\doibase 10.1103/PhysRevD.85.082003}
  {\bibfield  {journal} {\bibinfo  {journal} {Phys. Rev.}\ }\textbf {\bibinfo
  {volume} {D85}},\ \bibinfo {pages} {082003} (\bibinfo {year} {2012})},\
  \Eprint {http://arxiv.org/abs/1110.0530} {arXiv:1110.0530 [gr-qc]}
  \BibitemShut {NoStop}%
%%CITATION = ARXIV:1110.0530;%%
\bibitem [{\citenamefont {Ferrari}\ \emph {et~al.}(2001)\citenamefont
  {Ferrari}, \citenamefont {Pauri},\ and\ \citenamefont
  {Piazza}}]{Ferrari:2000ep}%
  \BibitemOpen
  \bibfield  {author} {\bibinfo {author} {\bibfnamefont {V.}~\bibnamefont
  {Ferrari}}, \bibinfo {author} {\bibfnamefont {M.}~\bibnamefont {Pauri}}, \
  and\ \bibinfo {author} {\bibfnamefont {F.}~\bibnamefont {Piazza}},\ }\href
  {\doibase 10.1103/PhysRevD.63.064009} {\bibfield  {journal} {\bibinfo
  {journal} {Phys. Rev.}\ }\textbf {\bibinfo {volume} {D63}},\ \bibinfo {pages}
  {064009} (\bibinfo {year} {2001})},\ \Eprint
  {http://arxiv.org/abs/gr-qc/0005125} {arXiv:gr-qc/0005125 [gr-qc]}
  \BibitemShut {NoStop}%
%%CITATION = GR-QC/0005125;%%
\bibitem [{\citenamefont {Molina}\ \emph {et~al.}(2010)\citenamefont {Molina},
  \citenamefont {Pani}, \citenamefont {Cardoso},\ and\ \citenamefont
  {Gualtieri}}]{Molina:2010fb}%
  \BibitemOpen
  \bibfield  {author} {\bibinfo {author} {\bibfnamefont {C.}~\bibnamefont
  {Molina}}, \bibinfo {author} {\bibfnamefont {P.}~\bibnamefont {Pani}},
  \bibinfo {author} {\bibfnamefont {V.}~\bibnamefont {Cardoso}}, \ and\
  \bibinfo {author} {\bibfnamefont {L.}~\bibnamefont {Gualtieri}},\ }\href
  {\doibase 10.1103/PhysRevD.81.124021} {\bibfield  {journal} {\bibinfo
  {journal} {Phys. Rev.}\ }\textbf {\bibinfo {volume} {D81}},\ \bibinfo {pages}
  {124021} (\bibinfo {year} {2010})},\ \Eprint {http://arxiv.org/abs/1004.4007}
  {arXiv:1004.4007 [gr-qc]} \BibitemShut {NoStop}%
%%CITATION = ARXIV:1004.4007;%%
\bibitem [{\citenamefont {Pani}\ and\ \citenamefont
  {Cardoso}(2009)}]{Pani:2009wy}%
  \BibitemOpen
  \bibfield  {author} {\bibinfo {author} {\bibfnamefont {P.}~\bibnamefont
  {Pani}}\ and\ \bibinfo {author} {\bibfnamefont {V.}~\bibnamefont {Cardoso}},\
  }\href {\doibase 10.1103/PhysRevD.79.084031} {\bibfield  {journal} {\bibinfo
  {journal} {Phys. Rev.}\ }\textbf {\bibinfo {volume} {D79}},\ \bibinfo {pages}
  {084031} (\bibinfo {year} {2009})},\ \Eprint {http://arxiv.org/abs/0902.1569}
  {arXiv:0902.1569 [gr-qc]} \BibitemShut {NoStop}%
%%CITATION = ARXIV:0902.1569;%%
\bibitem [{\citenamefont {Blazquez-Salcedo}\ \emph {et~al.}(2016)\citenamefont
  {Blazquez-Salcedo}, \citenamefont {Macedo}, \citenamefont {Cardoso},
  \citenamefont {Ferrari}, \citenamefont {Gualtieri}, \citenamefont {Khoo},
  \citenamefont {Kunz},\ and\ \citenamefont {Pani}}]{Blazquez-Salcedo:2016enn}%
  \BibitemOpen
  \bibfield  {author} {\bibinfo {author} {\bibfnamefont {J.~L.}\ \bibnamefont
  {Blazquez-Salcedo}}, \bibinfo {author} {\bibfnamefont {C.~F.~B.}\
  \bibnamefont {Macedo}}, \bibinfo {author} {\bibfnamefont {V.}~\bibnamefont
  {Cardoso}}, \bibinfo {author} {\bibfnamefont {V.}~\bibnamefont {Ferrari}},
  \bibinfo {author} {\bibfnamefont {L.}~\bibnamefont {Gualtieri}}, \bibinfo
  {author} {\bibfnamefont {F.~S.}\ \bibnamefont {Khoo}}, \bibinfo {author}
  {\bibfnamefont {J.}~\bibnamefont {Kunz}}, \ and\ \bibinfo {author}
  {\bibfnamefont {P.}~\bibnamefont {Pani}},\ }\href {\doibase
  10.1103/PhysRevD.94.104024} {\bibfield  {journal} {\bibinfo  {journal} {Phys.
  Rev.}\ }\textbf {\bibinfo {volume} {D94}},\ \bibinfo {pages} {104024}
  (\bibinfo {year} {2016})},\ \Eprint {http://arxiv.org/abs/1609.01286}
  {arXiv:1609.01286 [gr-qc]} \BibitemShut {NoStop}%
%%CITATION = ARXIV:1609.01286;%%
\bibitem [{\citenamefont {Blazquez-Salcedo}\ \emph {et~al.}(2017)\citenamefont
  {Blazquez-Salcedo}, \citenamefont {Khoo},\ and\ \citenamefont
  {Kunz}}]{Blazquez-Salcedo:2017txk}%
  \BibitemOpen
  \bibfield  {author} {\bibinfo {author} {\bibfnamefont {J.~L.}\ \bibnamefont
  {Blazquez-Salcedo}}, \bibinfo {author} {\bibfnamefont {F.~S.}\ \bibnamefont
  {Khoo}}, \ and\ \bibinfo {author} {\bibfnamefont {J.}~\bibnamefont {Kunz}},\
  }\href {\doibase 10.1103/PhysRevD.96.064008} {\bibfield  {journal} {\bibinfo
  {journal} {Phys. Rev.}\ }\textbf {\bibinfo {volume} {D96}},\ \bibinfo {pages}
  {064008} (\bibinfo {year} {2017})},\ \Eprint
  {http://arxiv.org/abs/1706.03262} {arXiv:1706.03262 [gr-qc]} \BibitemShut
  {NoStop}%
%%CITATION = ARXIV:1706.03262;%%
\bibitem [{\citenamefont {Brito}\ \emph
  {et~al.}(2013{\natexlab{a}})\citenamefont {Brito}, \citenamefont {Cardoso},\
  and\ \citenamefont {Pani}}]{Brito:2013wya}%
  \BibitemOpen
  \bibfield  {author} {\bibinfo {author} {\bibfnamefont {R.}~\bibnamefont
  {Brito}}, \bibinfo {author} {\bibfnamefont {V.}~\bibnamefont {Cardoso}}, \
  and\ \bibinfo {author} {\bibfnamefont {P.}~\bibnamefont {Pani}},\ }\href
  {\doibase 10.1103/PhysRevD.88.023514} {\bibfield  {journal} {\bibinfo
  {journal} {Phys. Rev.}\ }\textbf {\bibinfo {volume} {D88}},\ \bibinfo {pages}
  {023514} (\bibinfo {year} {2013}{\natexlab{a}})},\ \Eprint
  {http://arxiv.org/abs/1304.6725} {arXiv:1304.6725 [gr-qc]} \BibitemShut
  {NoStop}%
%%CITATION = ARXIV:1304.6725;%%
\bibitem [{\citenamefont {Brito}\ \emph
  {et~al.}(2013{\natexlab{b}})\citenamefont {Brito}, \citenamefont {Cardoso},\
  and\ \citenamefont {Pani}}]{Brito:2013yxa}%
  \BibitemOpen
  \bibfield  {author} {\bibinfo {author} {\bibfnamefont {R.}~\bibnamefont
  {Brito}}, \bibinfo {author} {\bibfnamefont {V.}~\bibnamefont {Cardoso}}, \
  and\ \bibinfo {author} {\bibfnamefont {P.}~\bibnamefont {Pani}},\ }\href
  {\doibase 10.1103/PhysRevD.87.124024} {\bibfield  {journal} {\bibinfo
  {journal} {Phys. Rev.}\ }\textbf {\bibinfo {volume} {D87}},\ \bibinfo {pages}
  {124024} (\bibinfo {year} {2013}{\natexlab{b}})},\ \Eprint
  {http://arxiv.org/abs/1306.0908} {arXiv:1306.0908 [gr-qc]} \BibitemShut
  {NoStop}%
%%CITATION = ARXIV:1306.0908;%%
\bibitem [{\citenamefont {Babichev}\ \emph {et~al.}(2016)\citenamefont
  {Babichev}, \citenamefont {Brito},\ and\ \citenamefont
  {Pani}}]{Babichev:2015zub}%
  \BibitemOpen
  \bibfield  {author} {\bibinfo {author} {\bibfnamefont {E.}~\bibnamefont
  {Babichev}}, \bibinfo {author} {\bibfnamefont {R.}~\bibnamefont {Brito}}, \
  and\ \bibinfo {author} {\bibfnamefont {P.}~\bibnamefont {Pani}},\ }\href
  {\doibase 10.1103/PhysRevD.93.044041} {\bibfield  {journal} {\bibinfo
  {journal} {Phys. Rev.}\ }\textbf {\bibinfo {volume} {D93}},\ \bibinfo {pages}
  {044041} (\bibinfo {year} {2016})},\ \Eprint
  {http://arxiv.org/abs/1512.04058} {arXiv:1512.04058 [gr-qc]} \BibitemShut
  {NoStop}%
%%CITATION = ARXIV:1512.04058;%%
\bibitem [{\citenamefont {Pani}\ \emph
  {et~al.}(2013{\natexlab{a}})\citenamefont {Pani}, \citenamefont {Berti},\
  and\ \citenamefont {Gualtieri}}]{Pani:2013ija}%
  \BibitemOpen
  \bibfield  {author} {\bibinfo {author} {\bibfnamefont {P.}~\bibnamefont
  {Pani}}, \bibinfo {author} {\bibfnamefont {E.}~\bibnamefont {Berti}}, \ and\
  \bibinfo {author} {\bibfnamefont {L.}~\bibnamefont {Gualtieri}},\ }\href
  {\doibase 10.1103/PhysRevLett.110.241103} {\bibfield  {journal} {\bibinfo
  {journal} {Phys. Rev. Lett.}\ }\textbf {\bibinfo {volume} {110}},\ \bibinfo
  {pages} {241103} (\bibinfo {year} {2013}{\natexlab{a}})},\ \Eprint
  {http://arxiv.org/abs/1304.1160} {arXiv:1304.1160 [gr-qc]} \BibitemShut
  {NoStop}%
%%CITATION = ARXIV:1304.1160;%%
\bibitem [{\citenamefont {Pani}\ \emph
  {et~al.}(2013{\natexlab{b}})\citenamefont {Pani}, \citenamefont {Berti},\
  and\ \citenamefont {Gualtieri}}]{Pani:2013wsa}%
  \BibitemOpen
  \bibfield  {author} {\bibinfo {author} {\bibfnamefont {P.}~\bibnamefont
  {Pani}}, \bibinfo {author} {\bibfnamefont {E.}~\bibnamefont {Berti}}, \ and\
  \bibinfo {author} {\bibfnamefont {L.}~\bibnamefont {Gualtieri}},\ }\href
  {\doibase 10.1103/PhysRevD.88.064048} {\bibfield  {journal} {\bibinfo
  {journal} {Phys. Rev.}\ }\textbf {\bibinfo {volume} {D88}},\ \bibinfo {pages}
  {064048} (\bibinfo {year} {2013}{\natexlab{b}})},\ \Eprint
  {http://arxiv.org/abs/1307.7315} {arXiv:1307.7315 [gr-qc]} \BibitemShut
  {NoStop}%
%%CITATION = ARXIV:1307.7315;%%
\bibitem [{\citenamefont {Mark}\ \emph {et~al.}(2015)\citenamefont {Mark},
  \citenamefont {Yang}, \citenamefont {Zimmerman},\ and\ \citenamefont
  {Chen}}]{Mark:2014aja}%
  \BibitemOpen
  \bibfield  {author} {\bibinfo {author} {\bibfnamefont {Z.}~\bibnamefont
  {Mark}}, \bibinfo {author} {\bibfnamefont {H.}~\bibnamefont {Yang}}, \bibinfo
  {author} {\bibfnamefont {A.}~\bibnamefont {Zimmerman}}, \ and\ \bibinfo
  {author} {\bibfnamefont {Y.}~\bibnamefont {Chen}},\ }\href {\doibase
  10.1103/PhysRevD.91.044025} {\bibfield  {journal} {\bibinfo  {journal} {Phys.
  Rev.}\ }\textbf {\bibinfo {volume} {D91}},\ \bibinfo {pages} {044025}
  (\bibinfo {year} {2015})},\ \Eprint {http://arxiv.org/abs/1409.5800}
  {arXiv:1409.5800 [gr-qc]} \BibitemShut {NoStop}%
%%CITATION = ARXIV:1409.5800;%%
\bibitem [{\citenamefont {Dias}\ \emph {et~al.}(2015)\citenamefont {Dias},
  \citenamefont {Godazgar},\ and\ \citenamefont {Santos}}]{Dias:2015wqa}%
  \BibitemOpen
  \bibfield  {author} {\bibinfo {author} {\bibfnamefont {O.~J.~C.}\
  \bibnamefont {Dias}}, \bibinfo {author} {\bibfnamefont {M.}~\bibnamefont
  {Godazgar}}, \ and\ \bibinfo {author} {\bibfnamefont {J.~E.}\ \bibnamefont
  {Santos}},\ }\href {\doibase 10.1103/PhysRevLett.114.151101} {\bibfield
  {journal} {\bibinfo  {journal} {Phys. Rev. Lett.}\ }\textbf {\bibinfo
  {volume} {114}},\ \bibinfo {pages} {151101} (\bibinfo {year} {2015})},\
  \Eprint {http://arxiv.org/abs/1501.04625} {arXiv:1501.04625 [gr-qc]}
  \BibitemShut {NoStop}%
%%CITATION = ARXIV:1501.04625;%%
\bibitem [{\citenamefont {Glampedakis}\ \emph {et~al.}(2017)\citenamefont
  {Glampedakis}, \citenamefont {Pappas}, \citenamefont {Silva},\ and\
  \citenamefont {Berti}}]{Glampedakis:2017dvb}%
  \BibitemOpen
  \bibfield  {author} {\bibinfo {author} {\bibfnamefont {K.}~\bibnamefont
  {Glampedakis}}, \bibinfo {author} {\bibfnamefont {G.}~\bibnamefont {Pappas}},
  \bibinfo {author} {\bibfnamefont {H.~O.}\ \bibnamefont {Silva}}, \ and\
  \bibinfo {author} {\bibfnamefont {E.}~\bibnamefont {Berti}},\ }\href
  {\doibase 10.1103/PhysRevD.96.064054} {\bibfield  {journal} {\bibinfo
  {journal} {Phys. Rev.}\ }\textbf {\bibinfo {volume} {D96}},\ \bibinfo {pages}
  {064054} (\bibinfo {year} {2017})},\ \Eprint
  {http://arxiv.org/abs/1706.07658} {arXiv:1706.07658 [gr-qc]} \BibitemShut
  {NoStop}%
%%CITATION = ARXIV:1706.07658;%%
\bibitem [{\citenamefont {Jai-akson}\ \emph {et~al.}(2017)\citenamefont
  {Jai-akson}, \citenamefont {Chatrabhuti}, \citenamefont {Evnin},\ and\
  \citenamefont {Lehner}}]{Jai-akson:2017ldo}%
  \BibitemOpen
  \bibfield  {author} {\bibinfo {author} {\bibfnamefont {P.}~\bibnamefont
  {Jai-akson}}, \bibinfo {author} {\bibfnamefont {A.}~\bibnamefont
  {Chatrabhuti}}, \bibinfo {author} {\bibfnamefont {O.}~\bibnamefont {Evnin}},
  \ and\ \bibinfo {author} {\bibfnamefont {L.}~\bibnamefont {Lehner}},\ }\href
  {\doibase 10.1103/PhysRevD.96.044031} {\bibfield  {journal} {\bibinfo
  {journal} {Phys. Rev.}\ }\textbf {\bibinfo {volume} {D96}},\ \bibinfo {pages}
  {044031} (\bibinfo {year} {2017})},\ \Eprint
  {http://arxiv.org/abs/1706.06519} {arXiv:1706.06519 [gr-qc]} \BibitemShut
  {NoStop}%
%%CITATION = ARXIV:1706.06519;%%
\bibitem [{\citenamefont {Glampedakis}\ and\ \citenamefont
  {Pappas}(2018)}]{Glampedakis:2017cgd}%
  \BibitemOpen
  \bibfield  {author} {\bibinfo {author} {\bibfnamefont {K.}~\bibnamefont
  {Glampedakis}}\ and\ \bibinfo {author} {\bibfnamefont {G.}~\bibnamefont
  {Pappas}},\ }\href {\doibase 10.1103/PhysRevD.97.041502} {\bibfield
  {journal} {\bibinfo  {journal} {Phys. Rev.}\ }\textbf {\bibinfo {volume}
  {D97}},\ \bibinfo {pages} {041502} (\bibinfo {year} {2018})},\ \Eprint
  {http://arxiv.org/abs/1710.02136} {arXiv:1710.02136 [gr-qc]} \BibitemShut
  {NoStop}%
%%CITATION = ARXIV:1710.02136;%%
\bibitem [{\citenamefont {Abbott}\ \emph
  {et~al.}(2016{\natexlab{f}})\citenamefont {Abbott} \emph
  {et~al.}}]{Abbott:2016nhf}%
  \BibitemOpen
  \bibfield  {author} {\bibinfo {author} {\bibfnamefont {B.~P.}\ \bibnamefont
  {Abbott}} \emph {et~al.} (\bibinfo {collaboration} {Virgo, LIGO
  Scientific}),\ }\href {\doibase 10.3847/2041-8205/833/1/L1} {\bibfield
  {journal} {\bibinfo  {journal} {Astrophys. J.}\ }\textbf {\bibinfo {volume}
  {833}},\ \bibinfo {pages} {L1} (\bibinfo {year} {2016}{\natexlab{f}})},\
  \Eprint {http://arxiv.org/abs/1602.03842} {arXiv:1602.03842 [astro-ph.HE]}
  \BibitemShut {NoStop}%
%%CITATION = ARXIV:1602.03842;%%
\bibitem [{\citenamefont {Purrer}(2016)}]{Purrer:2015tud}%
  \BibitemOpen
  \bibfield  {author} {\bibinfo {author} {\bibfnamefont {M.}~\bibnamefont
  {Purrer}},\ }\href {\doibase 10.1103/PhysRevD.93.064041} {\bibfield
  {journal} {\bibinfo  {journal} {Phys. Rev.}\ }\textbf {\bibinfo {volume}
  {D93}},\ \bibinfo {pages} {064041} (\bibinfo {year} {2016})},\ \Eprint
  {http://arxiv.org/abs/1512.02248} {arXiv:1512.02248 [gr-qc]} \BibitemShut
  {NoStop}%
%%CITATION = ARXIV:1512.02248;%%
\bibitem [{\citenamefont {Boh\'{e}}\ \emph {et~al.}(2017)\citenamefont
  {Boh\'{e}} \emph {et~al.}}]{Bohe:2016gbl}%
  \BibitemOpen
  \bibfield  {author} {\bibinfo {author} {\bibfnamefont {A.}~\bibnamefont
  {Boh\'{e}}} \emph {et~al.},\ }\href {\doibase 10.1103/PhysRevD.95.044028}
  {\bibfield  {journal} {\bibinfo  {journal} {Phys. Rev.}\ }\textbf {\bibinfo
  {volume} {D95}},\ \bibinfo {pages} {044028} (\bibinfo {year} {2017})},\
  \Eprint {http://arxiv.org/abs/1611.03703} {arXiv:1611.03703 [gr-qc]}
  \BibitemShut {NoStop}%
%%CITATION = ARXIV:1611.03703;%%
\end{thebibliography}%

\end{document}